\documentclass{LMCS}

\usepackage{url,amssymb,amsmath,amsthm,rotating,cite,hyperref}
\usepackage[curve]{xy}

\input{biernacka-biernacki-danvy-LMCS.mac}

\def\doi{1 (2:5) 2005}
\lmcsheading%
{\doi}
{39}
{}
{}
{Apr.~\phantom{0}1, 2005}
{Nov.~\phantom{0}8, 2005}
{}

\begin{document}

\title{An Operational Foundation for Delimited Continuations in the CPS Hierarchy}

\author[M.~Biernacka]{Ma{\l}gorzata Biernacka}
\author[D.~Biernacki]{Dariusz Biernacki}
\author[O.~Danvy]{Olivier Danvy}
\address{\protect\begin{tabular}[t]{@{}l}
         BRICS, Department of Computer Science, University of Aarhus
         \protect\\
         IT-parken, Aabogade 34, DK-8200 Aarhus N, Denmark
         \protect\end{tabular}}
\email{\{mbiernac,dabi,danvy\}@brics.dk}



\begin{abstract}

\noindent
  We present an abstract machine and a reduction semantics for the
  lambda-calculus extended with control operators that give access to
  delimited continuations in the CPS hierarchy. The abstract machine is
  derived from an evaluator in continuation-passing style (CPS); the
  reduction semantics (i.e., a small-step operational semantics with an
  explicit representation of evaluation contexts) is constructed from the
  abstract machine; and the control operators are the shift and reset
  family.
 
  We also present new applications of delimited continuations in the CPS
  hierarchy: finding list prefixes and normalization by evaluation for a
  hierarchical language of units and products.
\end{abstract}

\maketitle

\section{Introduction}
\label{sec:intro}

\noindent
The studies of delimited continuations can be classified in two
groups: those that use continu\-ation-passing style (CPS) and those
that rely on operational intuitions about control instead.
Of the latter, there is a large number proposing a variety of control
operators~\cite{%
Ariola-al:ICFP04,%
Felleisen:POPL88,%
Felleisen-al:TR87,%
Felleisen-al:LFP88,%
Gunter-al:FPCA95,%
Hieb-Dybvig:PPoPP90,%
Hieb-al:LaSC93,%
Moreau-Queinnec:PLILP94,%
Queinnec-Serpette:POPL91,%
Sitaram-Felleisen:LaSC90,%
Wadler:LaSC94-one%
} which have found applications in 
models of control, concurrency, and type-directed partial
evaluation~\cite{
Balat-al:POPL04,
Hieb-Dybvig:PPoPP90,
Sitaram-Felleisen:LFP90%
}.
Of the former, there is the work revolving around
the family of control operators shift and
reset~\cite{
Danvy-Filinski:DIKU89, 
Danvy-Filinski:LFP90,
Danvy-Filinski:MSCS92-one,
Danvy-Yang:ESOP99,
Filinski:POPL94,
Filinski:POPL99,
Kameyama:CSL04,%
Kameyama-Hasegawa:ICFP03,%
Murthy:CW92,%
Wadler:LaSC94-one%
}
which have found applications in 
non-deterministic programming,
code generation, 
partial evaluation,
normalization by evaluation,
computational monads, and
mobile computing~\cite{%
Asai:PEPM02,%
Asai:PEPM04,%
Balat-Danvy:GPCE02,%
Biernacki-Danvy:LOPSTR03,%
Danvy:POPL96,%
Danvy:PEPT98,%
Draves:ICFP97,%
Dybjer-Filinski:APPSEM00,%
Filinski:TLCA01,%
Gasbichler-Sperber:ICFP02,%
Grobauer-Yang:HOSC01,%
Helsen-Thiemann:ASIAN98,%
Kelsey-al:R5RS,%
Kiselyov:TR05,%
Lawall-Danvy:LFP94,
Shan:Scheme04,%
Sumii:Scheme00,%
Sumii-Kobayashi:HOSC01,%
Thiemann:JFP99%
}.

\myindent
The original motivation for shift and reset was a continuation-based
programming pattern involving several layers of continuations.  The
original specification of these operators relied both on a repeated
CPS transformation and on
an evaluator with several layers of continuations 
(as is obtained by repeatedly transforming a direct-style
evaluator into continuation-passing style).  Only subsequently have
shift and reset been specified operationally, by developing
operational analogues of a continuation semantics and of the CPS
transformation~\cite{Danvy-Yang:ESOP99}. 

\myindent
The goal of our work here is to establish a new operational foundation
for delimited continuations, using CPS as a guideline.  To this end,
we start with the original
evaluator for shift$_1$ and reset$_1$.
This
evaluator uses two layers of continuations: a continuation and a
meta-continuation.  We then defunctionalize it into an abstract
machine~\cite{Ager-al:PPDP03} and we construct the corresponding
reduction semantics~\cite{Felleisen:PhD}, as pioneered by Felleisen
and Friedman~\cite{Felleisen-Friedman:FDPC3}.
The development scales to shift$_n$ and reset$_n$.  It is reusable for
any control operators that are compatible with CPS, \ie, that can be
characterized with a (possibly iterated) CPS translation or with a
continuation-based evaluator.  It also pinpoints where operational
intuitions go beyond CPS.

\myindent
This article is structured as follows.
In Section~\ref{sec:arithmetic-expressions}, we review the enabling
technology of our work: Reynolds's defunctionalization, the
observation that a defunctionalized CPS program implements an abstract
machine, and the observation that Felleisen's evaluation contexts are
the defunctionalized continuations of a continuation-passing
evaluator; we demonstrate this enabling technology on a simple
example, arithmetic expressions.
In Section~\ref{sec:programming-with-shift-reset}, we illustrate the
use of shift and reset with the classic example of finding list
prefixes, using an ML-like programming language.
In Section~\ref{sec:derivation}, we then present our main result:
starting from the original
evaluator for shift and reset, we defunctionalize it into an abstract
machine;
we analyze this abstract machine and construct the corresponding
reduction semantics.
In Section~\ref{sec:up-in-the-hierarchy}, we extend 
this result
to the CPS hierarchy.
In Section~\ref{sec:programming-in-hierarchy}, we illustrate the CPS
hierarchy with a class of normalization functions for a hierarchical
language of
units and products.

\section{From evaluator to reduction semantics for arithmetic expressions}
\label{sec:arithmetic-expressions}

We demonstrate the derivation from an evaluator to a reduction semantics.
The derivation consists of the following steps:

\begin{enumerate}
  
\item we start from 
  an evaluator for a given language; if it is in direct style, we
  CPS-transform it;
  
\item we defunctionalize the CPS 
  evaluator, obtaining a value-based abstract machine;
  
\item we modify the abstract machine to make it term-based instead of
  value-based; in particular, if the
  evaluator uses an environment, then so does the
  corresponding value-based abstract machine, and in that case, making
  the machine term-based leads us to use substitutions rather than an
  environment;
  
\item we analyze the transitions of the
  term-based abstract machine to identify the evaluation strategy it
  implements and the set of reductions it performs; the result is a
  reduction semantics.

\end{enumerate}

\noindent
The first two steps are based on previous work on a functional
correspondence between evaluators and abstract
machines~\cite{Ager-al:PPDP03, Ager-al:IPL04, Ager-al:TCS05,
Biernacki-Danvy:LOPSTR03,
Danvy:IFL04}, which itself is based on Reynolds's seminal work on
definitional interpreters~\cite{Reynolds:HOSC98}.  The last two steps
follow the lines of Felleisen and Friedman's original work on a
reduction semantics for the call-by-value $\lambda$-calculus extended
with control operators~\cite{Felleisen-Friedman:FDPC3}.  The last step
has been studied further by Hardin, Maranget, and
Pagano~\cite{Hardin-al:JFP98} in the context of explicit
substitutions and by 
Biernacka, Danvy, and
Nielsen~\cite{Biernacka-Danvy:RS-05-15,Biernacka-Danvy:RS-05-22,Danvy-Nielsen:RS-04-26}.
 
In the rest of this section, our running example is the language of
arithmetic expressions, formed using natural numbers (the values) and
additions (the computations):
\[
\begin{array}{r@{\ }c@{\ }l}
\typeexp \ni \expr 
& ::= &
\synnum{m}
\Mid
\addition{\expr_1}{\expr_2}
\\
\end{array}
\]

\subsection{The starting point: an evaluator in direct style}

We define an evaluation function for arithmetic expressions by
structural induction on their syntax.  The resulting direct-style
evaluator is displayed in
Figure~\ref{fig:evaluator-arithmetic-expressions}.

\subsection{CPS transformation}

We CPS-transform the evaluator by naming
intermediate results,
sequentializing their computation, and introducing an extra functional
parameter, the
continuation~\cite{Danvy-Filinski:MSCS92-one,Plotkin:TCS75,Steele:MS}.
The resulting
continuation-passing evaluator is displayed in
Figure~\ref{fig:cps-evaluator-arithmetic-expressions}.

\begin{figure}[t]%
\hrule
\vspace{2mm}
\begin{itemize}
\itemsp{
Values:
$
\quad
\typeval \ni \val ::= \num{m} 
$}
\item
Evaluation function: 
\quad
$\evalar \;:\; \typeexp
\rightarrowp
\typeval$
\[
\begin{array}{rcl}
\namedone{\evalar}{\synnum{m}}
& = &
\num{m}
\\
\namedone{\evalar}{\addition{\expr_1}{\expr_2}}
& = &
\add{\namedone{\evalar}{\expr_1}}
    {\namedone{\evalar}{\expr_2}}

\end{array}
\]
\item
Main function:
\quad
$\main \;:\; \typeexp
  \rightarrowp 
 \typeval
 $
\[
\begin{array}{rcl}
\namedone{\main}{\expr}
& = &
\namedone{\evalar}{\expr}
\hspace{4.3cm}
\end{array}
\]
\end{itemize}
\caption{A direct-style evaluator for arithmetic expressions}
\label{fig:evaluator-arithmetic-expressions}
\vspace{2mm}
\hrule
\end{figure}

\begin{figure}
\hrule
\vspace{2mm}
\begin{itemize}
\itemsp{
Values:
$
\quad
\typeval \ni \val ::= \num{m}
$}

\itemsp{
Continuations:
\quad
$
\typecont\equals\typeval\rightarrowp\typeval
$} 

\item
Evaluation function:
\quad
$
\oftype{\evalar}{\typeexp\timesp\typecont\rightarrowp\typeval}
$
\[
\begin{array}{rcl}
\namedtwo{\evalar}{\synnum{m}}{k}
& = &
\applic{k}{\num{m}}
\\
\namedtwo{\evalar}{\addition{\expr_1}{\expr_2}}{k}
& = &
\namedtwo{\evalar}
         {\expr_1}
         {\lamabs{\num{m_1}}{\namedtwo{\evalar}
                                      {\expr_2}
                                      {\lamabs{\num{m_2}}{\applic{k}{(\add{m_1}{m_2})}}}}}
\end{array}
\]

\item
Main function:
\quad
$\main \;:\; \typeexp
\rightarrowp 
\typeval
$
\[
\begin{array}{rcl}
\namedone{\main}{\expr}
& = &
\namedtwo{\evalar}{\expr}{\lamabs{\val}{\val}}
\hspace{2.2cm}
\end{array}
\]
\end{itemize}
\caption{A
continuation-passing evaluator for arithmetic expressions}
\label{fig:cps-evaluator-arithmetic-expressions}
\vspace{2mm}
\hrule
\end{figure}

\begin{figure}[ht!]%
\hrule
\vspace{2mm}
\begin{itemize}
\itemsp{
Values:
$
\quad
\typeval \ni \val ::= \num{m}
$}

\itemsp{
Defunctionalized continuations:
$
\quad
\typecont \ni \ka ::=  
\contzeroo
\Mid
\contonee{\expr}{\ka}
\Mid
\conttwoo{\val}{\ka}
$}
\item
Functions
$\oftype{\evalar}{\typeexp\timesp\typecont\rightarrowp\typeval}$
and 
$\oftype{\apply}{\typecont\timesp\typeval\rightarrowp\typeval}$:
\vspace{-0.1cm}
\[
\begin{array}{rcl}
\\
\namedtwo{\evalar}{\synnum{m}}{\ka}
& = &
\namedtwo{\apply}{\ka}{\num{m}}
\\
\namedtwo{\evalar}{\addition{\expr_1}{\expr_2}}{\ka}
& = &
\namedtwo{\evalar}{\expr_1}{\contonee{\expr_2}{\ka}}
\\
\\
\namedtwo{\apply}{\contzeroo}{\val}
& = &
\val
\\
\namedtwo{\apply}{\contonee{\expr_2}{\ka}}{\val_1}
& = &
\namedtwo{\evalar}{\expr_2}{\conttwoo{\val_1}{\ka}}
\\
\namedtwo{\apply}{\conttwoo{\num{m_1}}{\ka}}{\num{m_2}}
& = &
\namedtwo{\apply}{\ka}{\add{\num{m_1}}{\num{m_2}}}
\end{array}
\]
\item
Main function:
$\oftype{\main}{\typeexp\rightarrowp\typeval}$
\[
\begin{array}{rcl}
\namedone{\main}{\expr}
& = &
\namedtwo{\evalar}{\expr}{\contzeroo}
\end{array}
\]
\end{itemize}
\caption{A defunctionalized
continuation-passing evaluator for arithmetic expressions}
\label{fig:defunctionalized-cps-evaluator-arithmetic-expressions}
\vspace{2mm}
\hrule
\end{figure}

\subsection{Defunctionalization}

The generalization of closure conversion~\cite{Landin:CJ64} to
defunctionalization is due to Reynolds~\cite{Reynolds:HOSC98}.  The
goal is to represent a functional value with a first-order data
structure.  The means is to partition the function space into a
first-order sum where each summand corresponds to a lambda-abstraction
in the
program.  In a defunctionalized
program, function
introduction is thus represented as an injection, and function
elimination as a call to a first-order apply function implementing a
case dispatch.  In an ML-like functional language, sums are
represented as data types, injections as data-type constructors, and
apply functions are defined by case over the corresponding data
types~\cite{Danvy-Nielsen:PPDP01}.

Here, we defunctionalize the continuation of the continuation-passing
evaluator in Figure~\ref{fig:cps-evaluator-arithmetic-expressions}.  We
thus need to define a first-order algebraic data type and its apply
function.  To this end, we enumerate the lambda-abstractions that give
rise to the inhabitants of this function space; there are three: the
initial continuation in $\main$ and the two continuations in $\evalar$.
The initial continuation is closed, and therefore the corresponding
algebraic constructor is nullary.  The two other continuations have two
free variables, and therefore the corresponding constructors are binary.
As for the apply function, it interprets the algebraic constructors.  The
resulting defunctionalized evaluator is displayed in
Figure~\ref{fig:defunctionalized-cps-evaluator-arithmetic-expressions}.

\subsection{Abstract machines as defunctionalized continuation-passing programs}

Elsewhere~\cite{Ager-al:PPDP03,
Danvy:IFL04}, we have observed that
a defunctionalized continuation-passing program implements an abstract
machine: each configuration is the name of a function together with
its arguments, and each function clause represents a transition.  (As
a corollary, we have also observed that the defunctionalized
continuation of an evaluator forms what is known as an `evaluation
context'~\cite{Danvy:CW04,Danvy-Nielsen:PPDP01,Felleisen-Friedman:FDPC3}.)

Indeed Plotkin's Indifference Theorem~\cite{Plotkin:TCS75} states that
continuation-passing programs are independent of their evaluation order.
In Reynolds's words~\cite{Reynolds:HOSC98}, all the subterms in
applications are `trivial'; and in Moggi's words~\cite{Moggi:IaC91},
these subterms are values and not computations.  Furthermore,
continuation-passing programs are tail recursive~\cite{Steele:MS}.
Therefore, since in a continuation-passing program all calls are tail
calls and all subcomputations are elementary, a defunctionalized
continuation-passing program implements a transition
system~\cite{Plotkin:TR81}, \ie, an abstract machine.

We thus reformat
Figure~\ref{fig:defunctionalized-cps-evaluator-arithmetic-expressions}
into Figure~\ref{fig:abstract-machine-arithmetic-expressions}.  The
correctness of the abstract machine with respect to the initial
evaluator
follows from the correctness of CPS transformation and of
defunctionalization.

\begin{figure}[ht]%
\hrule
\vspace{2mm}
\begin{itemize}
\itemsp{
Values:
$
\quad
\val ::= \num{m}
$}

\itemsp{
Evaluation contexts:
\quad
$\cnt
 ::= 
\mtcont
\Mid
\cntexp{\expr}
\Mid
\cntval{\val}
$}
\item
Initial transition, transition rules, and final transition:
\[
\begin{machinetable}

\spacearound
\transition{\expr}
           {\namedpair{\evalconfname}
                      {\expr}
                      {\mtcont}
                    }

\vspace{3mm}

\spacearound
\transition{\namedpair{\evalconfname}
                      {\synnum{m}}
                      {\cnt}
                    }
           {\namedpair{\applyconfname}
                      {\cnt}
                      {\num{m}}
                    }

\spacearound
\hspace{0.2cm}
\transition{\namedpair{\evalconfname}
                      {\addition{\expr_1}{\expr_2}}
                      {\cnt}
                    }
           {\namedpair{\evalconfname}
                      {\expr_1}
                      {\cntexp{\expr_2}}
                    }

\vspace{3mm}

\spacearound
\transition{\namedpair{\applyconfname}
                      {\cntexp{\expr_2}}
                      {\val_1}
                    }
           {\namedpair{\evalconfname}
                      {\expr_2}
                      {\cntval{\val_1}}
                    }

\spacearound
\transition{\namedpair{\applyconfname}
                      {\cntval{\num{m_1}}}
                      {\num{m_2}}
                    }
           {\namedpair{\applyconfname}
                      {\cnt}
                      {\additionsem{\num{m_1}}{\num{m_2}}}
                    }
                    
\vspace{3mm}

\spacearound
\transition{\namedpair{\applyconfname}
                      {\mtcont}
                      {\val}
                    }
           {\val}

\end{machinetable}
\]
\end{itemize}
\vspace{-0.2cm}
\caption{A value-based abstract machine for evaluating arithmetic expressions}
\label{fig:abstract-machine-arithmetic-expressions}
\vspace{2mm}
\hrule
\end{figure}

\begin{figure}%
\hrule
\vspace{2mm}
\begin{itemize}
\itemsp{
Expressions and values:
$
\begin{array}[t]{r@{\ }c@{\ }l}
\expr
& ::= &
\val
\Mid
\addition{\expr_1}{\expr_2}
\\
\val
& ::= &
\synnum{m}
\end{array}
$}
\itemsp{
Evaluation contexts:
\quad
$\cnt
 ::= 
\mtcont
\Mid
\cntexp{\expr}
\Mid
\cntval{\val}
$}
\item
Initial transition, transition rules, and final transition:
\vspace{0.2cm}
\[
\begin{machinetable}

\vspace{3mm}
\spacearound
\transition{\expr}
           {\namedpair{\evalconfname}
                      {\expr}
                      {\mtcont}
                    }

\spacearound
\transition{\namedpair{\evalconfname}
                      {\synnum{m}}
                      {\cnt}
                    }
           {\namedpair{\applyconfname}
                      {\cnt}
                      {\synnum{m}}
                    }

\spacearound
\vspace{3mm}
\transition{\namedpair{\evalconfname}
                      {\addition{\expr_1}{\expr_2}}
                      {\cnt}
                    }
           {\namedpair{\evalconfname}
                      {\expr_1}
                      {\cntexp{\expr_2}}
                    }

\spacearound
\transition{\namedpair{\applyconfname}
                      {\cntexp{\expr_2}}
                      {\val_1}
                    }
           {\namedpair{\evalconfname}
                      {\expr_2}
                      {\cntval{\val_1}}
                    }

\vspace{3mm}
\spacearound
\transition{\namedpair{\applyconfname}
                      {\cntval{\synnum{m_1}}}
                      {\synnum{m_2}}
                    }
           {\namedpair{\applyconfname}
                      {\cnt}
                      {\cornered{\additionsem{\num{m_1}}{\num{m_2}}}}
                    }

\spacearound
\transition{\namedpair{\applyconfname}
                      {\mtcont}
                      {\val}
                    }
           {\val}

\end{machinetable}
\]
\end{itemize}
\caption{A term-based abstract machine for processing arithmetic expressions}
\label{fig:abstract-machine-arithmetic-expressions-syntax}
\vspace{2mm}
\hrule
\end{figure}

\subsection{From value-based abstract machine to term-based abstract machine}

We observe that the domain of expressible values in
Figure~\ref{fig:abstract-machine-arithmetic-expressions} can be embedded
in the syntactic domain of
expressions.  We therefore adapt the
abstract machine to work on terms rather than on values.  The result is
displayed in
Figure~\ref{fig:abstract-machine-arithmetic-expressions-syntax}; it is a
syntactic theory~\cite{Felleisen:PhD}.

\subsection{From term-based abstract machine to reduction semantics}
\label{subsec:from-term-based-abstract-machine-to-reduction-semantics}

The method of deriving a reduction semantics from an abstract machine was
introduced by Felleisen and Friedman~\cite{Felleisen-Friedman:FDPC3} to
give a reduction semantics for control operators.
Let us demonstrate it.

We analyze the transitions of the abstract machine in
Figure~\ref{fig:abstract-machine-arithmetic-expressions-syntax}.
The second component of $\mathit{eval}$-transitions---the stack
representing ``the rest of the computation''---has already been
identified as the evaluation context of the currently processed
expression.  We thus read a configuration
$\namedpair{\evalconfname}{\expr}{\cnt}$ as a decomposition of some
expression into a sub-expression $\expr$ and an evaluation context
$\cnt$.

Next, we identify the reduction and decomposition rules in the
transitions of the machine. Since a configuration can be read as a
decomposition, we compare the left-hand side and the right-hand side
of each transition. If they represent the same expression, then the
given transition defines a decomposition (\ie, it searches for the
next redex according to some evaluation strategy); otherwise we have
found a redex. Moreover, reading the
decomposition rules from right to left defines a `plug' function that
reconstructs an expression from its decomposition.

Here the decomposition function as read off the abstract machine is
total. In general, however, it may be undefined for stuck terms;
one can then extend it straightforwardly into a total function that
decomposes
a term into a context and a \emph{potential redex}, \ie, an \emph{actual
  redex} (as read off the machine), or a \emph{stuck redex}.

In
this simple example there is only one reduction rule.  This rule performs
the addition of natural numbers:
\[
\begin{array}{lrcl}
\redadd
&
\decomp{\cnt}{\addition{\synnum{m_1}}{\synnum{m_2}}}
&
\red
&
\decomp{\cnt}{\cornered{\additionsem{\num{m_1}}{\num{m_2}}}}
\end{array}
\]
\noindent
The remaining transitions decompose an expression according to the
left-to-right strategy.

\subsection{From reduction semantics to term-based abstract machine}

In
Section~\ref{subsec:from-term-based-abstract-machine-to-reduction-semantics},
we have constructed the reduction semantics corresponding to the abstract
machine of
Figure~\ref{fig:abstract-machine-arithmetic-expressions-syntax}, as
pioneered by Felleisen and
Friedman~\cite{Felleisen-Flatt:LN,Felleisen-Friedman:FDPC3}.  Over the
last few
years~\cite{Biernacka-Danvy:RS-05-15,Biernacka-Danvy:RS-05-22,Danvy:WRS04,Danvy-Nielsen:RS-04-26},
Biernacka, Danvy, and
Nielsen have studied the converse transformation and systematized the
construction of an abstract machine from a reduction semantics.  The
main idea is to short-cut the decompose-contract-plug loop, in the
definition of evaluation as the transitive closure of one-step reduction,
into a refocus-contract loop.  The refocus function is constructed as an
efficient (\ie, deforested) composition of plug and decompose that maps a
term and a context either to a value or to a redex and a context.  The
result is a `pre-abstract machine' computing the transitive closure of
the refocus function.  This pre-abstract machine can then be simplified
into
an eval/apply abstract machine.

It is simple to verify that using refocusing, one
can go from the reduction semantics of
Section~\ref{subsec:from-term-based-abstract-machine-to-reduction-semantics}
to the eval/apply abstract machine of
Figure~\ref{fig:abstract-machine-arithmetic-expressions-syntax}.

\subsection{Summary and conclusion}

We have demonstrated how to derive an abstract machine out of
an evaluator, and how to construct the corresponding reduction semantics
out of this abstract machine.  In Section~\ref{sec:derivation}, we apply
this derivation and this construction to the first level of the CPS
hierarchy, and in Section~\ref{sec:up-in-the-hierarchy}, we apply them to
an arbitrary level of the CPS hierarchy.  But first, let us illustrate
how to program with delimited continuations.

\section{Programming with delimited continuations}
\label{sec:programming-with-shift-reset}

We
present two examples of programming with delimited continuations.
Given a list $\synvar{xs}$ and a predicate $\synvar{p}$, we want
\begin{enumerate}
  
\item to find the first prefix of $\synvar{xs}$ whose last element
  satisfies $\synvar{p}$, and
  
\item to find all such prefixes of $\synvar{xs}$.

\end{enumerate}

\noindent
For example, given the predicate
$\synlam{m}{m > 2}$ and the list $[0,3,1,4,2,5]$, the first prefix is
$[0,3]$ and the list of all the prefixes is
$[[0,3],[0,3,1,4],[0,3,1,4,2,5]]$.

In Section~\ref{subsec:finding-prefixes-by-accumulating-lists}, we start
with a simple solution that uses a first-order accumulator.  This simple
solution is in defunctionalized form.  In
Section~\ref{subsec:finding-prefixes-by-accumulating-list-constructors},
we present its higher-order counterpart, which uses a functional
accumulator.  This functional accumulator acts as a delimited
continuation.  In Section~\ref{subsec:finding-prefixes-in-direct-style},
we present its direct-style counterpart (which uses shift and reset) and
in Section~\ref{subsec:finding-prefixes-in-continuation-passing-style},
we present its continuation-passing counterpart (which uses two layers of
continuations).  In Section~\ref{subsec:the-CPS-hierarchy}, we introduce
the CPS hierarchy informally.  We then mention a typing issue in
Section~\ref{subsec:a-note-about-typing} and review related work in
Section~\ref{subsection:related-work}.

\subsection{Finding prefixes by accumulating lists}
\label{subsec:finding-prefixes-by-accumulating-lists}

A simple solution is to accumulate the prefix of the given list in
reverse order while traversing this list and testing each of its
elements:
\begin{itemize}
  
\item if no element satisfies the predicate, there is no prefix and
  the result is the empty list;
  
\item otherwise, the prefix is the reverse of the accumulator.

\end{itemize}

\vspace{-6mm}

\begin{eqnarray*}
   \synapp{\synvar{find\_first\_prefix\_a}}
          {\synpair{\synvar{p}}{\synvar{xs}}}
   &
   \stackrel{\mathrm{def}}{=}
   &
   \vsynletrectwoandtwo{\synapp{\synvar{visit}}
                               {\synpair{\synnil}
                                        {\synvar{a}}}}
                       {\synnil}
                       {\synapp{\synvar{visit}}
                               {\synpair{\syncons{\synvar{x}}
                                                 {\synvar{xs}}}
                                        {\synvar{a}}}}
                       {\vsynlet{\synvar{a'}}
                                {\syncons{\synvar{x}}{\synvar{a}}}
                                {\vsynif{\synapp{\synvar{p}}{\synvar{x}}}
                                        {\synapp{\synvar{reverse}}{\synpair{\synvar{a'}}
                                                                           {\synnil}}}
                                        {\synapp{\synvar{visit}}{\synpair{\synvar{xs}}{\synvar{a'}}}}}}
                       {\synapp{\synvar{reverse}}
                               {\synpair{\synnil}
                                        {\synvar{xs}}}}
                       {\synvar{xs}}
                       {\synapp{\synvar{reverse}}
                               {\synpair{\syncons{\synvar{x}}{\synvar{a}}}
                                        {\synvar{xs}}}}
                       {\synapp{\synvar{reverse}}
                               {\synpair{\synvar{a}}
                                        {\syncons{\synvar{x}}{\synvar{xs}}}}}
                       {\synapp{\synvar{visit}}{\synpair{\synvar{xs}}{\synnil}}}
\\[2ex]
   \synapp{\synvar{find\_all\_prefixes\_a}}
          {\synpair{\synvar{p}}{\synvar{xs}}}
   &
   \stackrel{\mathrm{def}}{=}
   &
   \vsynletrectwoandtwo{\synapp{\synvar{visit}}
                               {\synpair{\synnil}
                                        {\synvar{a}}}}
                       {\synnil}
                       {\synapp{\synvar{visit}}
                               {\synpair{\syncons{\synvar{x}}
                                                 {\synvar{xs}}}
                                        {\synvar{a}}}}
                       {\vsynlet{\synvar{a'}}
                                {\syncons{\synvar{x}}{\synvar{a}}}
                                {\vsynif{\synapp{\synvar{p}}{\synvar{x}}}
                                        {\syncons{\synappp{\synvar{reverse}}{\synpair{\synvar{a'}}
                                                                                     {\synnil}}}
                                                 {\synappp{\synvar{visit}}{\synpair{\synvar{xs}}{\synvar{a'}}}}}
                                        {\synapp{\synvar{visit}}{\synpair{\synvar{xs}}{\synvar{a'}}}}}}
                       {\synapp{\synvar{reverse}}
                               {\synpair{\synnil}
                                        {\synvar{xs}}}}
                       {\synvar{xs}}
                       {\synapp{\synvar{reverse}}
                               {\synpair{\syncons{\synvar{x}}{\synvar{a}}}
                                        {\synvar{xs}}}}
                       {\synapp{\synvar{reverse}}
                               {\synpair{\synvar{a}}
                                        {\syncons{\synvar{x}}{\synvar{xs}}}}}
                       {\synapp{\synvar{visit}}{\synpair{\synvar{xs}}{\synnil}}}
\end{eqnarray*}

\vspace{1mm}

\noindent
To find the first prefix, one stops as soon as a satisfactory list
element is found.  To list all the prefixes, one continues the
traversal, adding the current prefix to the list of the remaining
prefixes.

We observe that the two solutions are in defunctionalized
form~\cite{Danvy-Nielsen:PPDP01,Reynolds:HOSC98}: the accumulator has
the data type of a defunctionalized function and $\synvar{reverse}$ is
its apply function.  We present its higher-order counterpart
next~\cite{Hughes:IPL86}.

\subsection{Finding prefixes by accumulating list constructors}
\label{subsec:finding-prefixes-by-accumulating-list-constructors}

Instead of accumulating the prefix in reverse order while traversing
the given list, we accumulate a function constructing the prefix:
\begin{itemize}
  
\item if no element satisfies the predicate, the result is the empty
  list;

\item otherwise, we apply the functional accumulator to construct the
  prefix.
\end{itemize}
\begin{eqnarray*}
\vspace{-6mm}
   \synapp{\synvar{find\_first\_prefix\_c}_1}
          {\synpair{\synvar{p}}{\synvar{xs}}}
   &
   \stackrel{\mathrm{def}}{=}
   &
   \vsynletrectwo{\synapp{\synvar{visit}}
                         {\synpair{\synnil}
                                  {\synvar{k}}}}
                 {\synnil}
                 {\synapp{\synvar{visit}}
                         {\synpair{\syncons{\synvar{x}}
                                           {\synvar{xs}}}
                                  {\synvar{k}}}}
                 {\vsynlet{\synvar{k'}}
                          {\synlam{\synvar{vs}}
                                  {\synapp{\synvar{k}}
                                          {\synconsp{\synvar{x}}{\synvar{vs}}}}}
                          {\vsynif{\synapp{\synvar{p}}{\synvar{x}}}
                                  {\synapp{\synvar{k'}}{\synnil}}
                                  {\synapp{\synvar{visit}}{\synpair{\synvar{xs}}{\synvar{k'}}}}}}
                 {\synapp{\synvar{visit}}{\synpair{\synvar{xs}}{\synlam{\synvar{vs}}{\synvar{vs}}}}}
\\[2ex]
   \synapp{\synvar{find\_all\_prefixes\_c}_1}
          {\synpair{\synvar{p}}{\synvar{xs}}}
   &
   \stackrel{\mathrm{def}}{=}
   &
   \vsynletrectwo{\synapp{\synvar{visit}}
                         {\synpair{\synnil}
                                  {\synvar{k}}}}
                 {\synnil}
                 {\synapp{\synvar{visit}}
                         {\synpair{\syncons{\synvar{x}}
                                           {\synvar{xs}}}
                                  {\synvar{k}}}}
                 {\vsynlet{\synvar{k'}}
                          {\synlam{\synvar{vs}}
                                  {\synapp{\synvar{k}}
                                          {\synconsp{\synvar{x}}{\synvar{vs}}}}}
                          {\vsynif{\synapp{\synvar{p}}{\synvar{x}}}
                                  {\syncons{\synappp{\synvar{k'}}{\synnil}}
                                           {\synappp{\synvar{visit}}{\synpair{\synvar{xs}}{\synvar{k'}}}}}
                                  {\synapp{\synvar{visit}}{\synpair{\synvar{xs}}{\synvar{k'}}}}}}
                 {\synapp{\synvar{visit}}{\synpair{\synvar{xs}}{\synlam{\synvar{vs}}{\synvar{vs}}}}}
\end{eqnarray*}

\vspace{1mm}

\noindent
To find the first prefix, one applies the functional accumulator as
soon as a satisfactory list element is found.  To list all such
prefixes, one continues the traversal, adding the current prefix to
the list of the remaining prefixes.

Defunctionalizing these two definitions yields the two definitions of
Section~\ref{subsec:finding-prefixes-by-accumulating-lists}.

The functional accumulator is a delimited continuation:

\begin{itemize}

\item
In $\synvar{find\_first\_prefix\_c}_1$, $\synvar{visit}$ is written in
CPS since all calls are tail calls and all sub-computations are
elementary.  The continuation is initialized in the initial call to
$\synvar{visit}$, discarded in the base case, extended in the
induction case, and used if a satisfactory prefix is found.

\item
In $\synvar{find\_all\_prefixes\_c}_1$, $\synvar{visit}$ is almost
written in
CPS except that the continuation is composed if a satisfactory prefix
is found: it is  
used twice---once where it is applied to the empty list to construct a
prefix, and once 
in the visit of the rest of the list to construct a list of prefixes;
this prefix is then prepended to 
this list of prefixes.

\end{itemize}

\noindent
These continuation-based programming patterns (initializing a
continuation, not using it, or using it more than once as if it were a
composable function) have motivated the control operators shift and
reset~\cite{Danvy-Filinski:LFP90,Danvy-Filinski:MSCS92-one}.  Using
them, in the next section, we write $\synvar{visit}$ in direct style.

\subsection{Finding prefixes in direct style}
\label{subsec:finding-prefixes-in-direct-style}

The two following local functions are the direct-style counterpart of
the two local functions in
Section~\ref{subsec:finding-prefixes-by-accumulating-list-constructors}:

\begin{eqnarray*}
   &&
   \\[-1cm]
   \synapp{\synvar{find\_first\_prefix\_c}_0}
          {\synpair{\synvar{p}}{\synvar{xs}}}
   &
   \stackrel{\mathrm{def}}{=}
   &
   \vsynletrectwo{\synapp{\synvar{visit}}
                         {\synnil}}
                 {\synshift{k}{\synnil}}
                 {\synapp{\synvar{visit}}
                         {\synconsp{\synvar{x}}
                                   {\synvar{xs}}}}
                 {\syncons{\synvar{x}}
                          {\hsynifp{\synapp{\synvar{p}}{\synvar{x}}}
                                   {\synnil}
                                   {\synapp{\synvar{visit}}{\synvar{xs}}}}}
                 {\synreset{\synapp{\synvar{visit}}{\synvar{xs}}}}
\end{eqnarray*}
\begin{eqnarray*}
   \synapp{\synvar{find\_all\_prefixes\_c}_0}
          {\synpair{\synvar{p}}{\synvar{xs}}}
   &
   \stackrel{\mathrm{def}}{=}
   &
   \vsynletrectwo{\synapp{\synvar{visit}}
                         {\synnil}}
                 {\synshift{k}{\synnil}}
                 {\synapp{\synvar{visit}}
                         {\synconsp{\synvar{x}}
                                   {\synvar{xs}}}}
                 {\syncons{\synvar{x}}
                          {\vsynif{\synapp{\synvar{p}}{\synvar{x}}}
                                  {\synshift{k'}
                                            {\syncons{\synreset{\synapp{\synvar{k'}}
                                                                       {\synnil}}}
                                                     {\synreset{\synapp{k'}
                                                                       {\synappp{\synvar{visit}}
                                                                                {\synvar{xs}}}}}}}
                                  {\synapp{\synvar{visit}}{\synvar{xs}}}}}
                 {\synreset{\synapp{\synvar{visit}}{\synvar{xs}}}}
\end{eqnarray*}

\noindent
In both cases, $\synvar{visit}$ is in direct style, \ie, it is not passed
any continuation.  The initial calls to $\synvar{visit}$ are enclosed in
the control delimiter reset (noted $\rawreset$ for conciseness).  In the
base cases, the current (delimited) continuation is captured with the
control operator shift (noted $\rawshift$), which has the effect of
emptying the (delimited) context; this captured continuation is bound to
an identifier $k$, which is not used; $\synnil$ is then returned in the
emptied context.  In the induction case of
$\synvar{find\_all\_prefixes\_c}_0$, if the predicate is satisfied,
$\synvar{visit}$ captures the current continuation and applies it
twice---once to the empty list to construct a prefix, and once to the
result of visiting the rest of the list to construct a list of prefixes;
this prefix is then prepended to the list of prefixes.

CPS-transforming these two local
functions yields the two definitions of
Section~\ref{subsec:finding-prefixes-by-accumulating-list-constructors}~\cite{Danvy-Filinski:MSCS92-one}.

\subsection{Finding prefixes in continuation-passing style}
\label{subsec:finding-prefixes-in-continuation-passing-style}

The two following local functions are the continuation-passing
counterpart of the two local functions in
Section~\ref{subsec:finding-prefixes-by-accumulating-list-constructors}:
\[
   \begin{array}{@{\hspace{-1mm}}r@{\ }c@{\ }l@{\hspace{-1mm}}}
   \synapp{\synvar{find\_first\_prefix\_c}_2}
          {\synpair{\synvar{p}}{\synvar{xs}}}
   &
   \stackrel{\mathrm{def}}{=}
   &
   \vsynletrectwo{\synapp{\synvar{visit}}
                         {\syntriple{\synnil}
                                    {\synvar{k}_1}
                                    {\synvar{k}_2}}}
                 {\synapp{\synvar{k}_2}{\synnil}}
                 {\synapp{\synvar{visit}}
                         {\syntriple{\syncons{\synvar{x}}
                                             {\synvar{xs}}}
                                    {\synvar{k}_1}
                                    {\synvar{k}_2}}}
                 {\vsynlet{\synvar{k}_1'}
                          {\synlam{\synpair{\synvar{vs}}{\synvar{k}_2'}}
                                  {\synapp{\synvar{k}_1}
                                          {\synpair{\syncons{\synvar{x}}{\synvar{vs}}}
                                                   {\synvar{k}_2'}}}}
                          {\vsynif{\synapp{\synvar{p}}{\synvar{x}}}
                                  {\synapp{\synvar{k}_1'}
                                          {\synpair{\synnil}
                                                   {\synvar{k}_2}}}
                                  {\synapp{\synvar{visit}}{\syntriple{\synvar{xs}}{\synvar{k}_1'}{\synvar{k}_2}}}}}
                 {\synapp{\synvar{visit}}
                         {\syntriple{\synvar{xs}}
                                    {\synlam{\synpair{\synvar{vs}}{\synvar{k}_2}}
                                            {\synapp{\synvar{k}_2}{\synvar{vs}}}}
                                    {\synlam{\synvar{vs}}{\synvar{vs}}}}}
\\[3.7cm]
   \synapp{\synvar{find\_all\_prefixes\_c}_2}
          {\synpair{\synvar{p}}{\synvar{xs}}}
   &
   \stackrel{\mathrm{def}}{=}
   &
   \vsynletrectwo{\synapp{\synvar{visit}}
                         {\syntriple{\synnil}
                                    {\synvar{k}_1}
                                    {\synvar{k}_2}}}
                 {\synapp{\synvar{k}_2}{\synnil}}
                 {\synapp{\synvar{visit}}
                         {\syntriple{\syncons{\synvar{x}}
                                             {\synvar{xs}}}
                                    {\synvar{k}_1}
                                    {\synvar{k}_2}}}
                 {\vsynlet{\synvar{k}_1'}
                          {\synlam{\synpair{\synvar{vs}}{\synvar{k}_2'}}
                                  {\synapp{\synvar{k}_1}
                                          {\synpair{\syncons{\synvar{x}}{\synvar{vs}}}
                                                   {\synvar{k}_2'}}}}
                          {\vsynif{\synapp{\synvar{p}}{\synvar{x}}}
                                  {\synapp{\synvar{k}_1'}
                                          {\synpair{\synnil}
                                                   {\synlam{\synvar{vs}}
                                                           {\synapp{\synvar{visit}}
                                                                   {\syntriple{\synvar{xs}}
                                                                              {\synvar{k}_1'}
                                                                              {\synlam{\synvar{vss}}
                                                                                      {\synapp{\synvar{k}_2}
                                                                                              {\synconsp{\synvar{vs}}
                                                                                                       {\synvar{vss}}}}}}}}}}
                                  {\synapp{\synvar{visit}}{\syntriple{\synvar{xs}}{\synvar{k}_1'}{\synvar{k}_2}}}}}
                 {\synapp{\synvar{visit}}
                         {\syntriple{\synvar{xs}}
                                    {\synlam{\synpair{\synvar{vs}}{\synvar{k}_2}}
                                            {\synapp{\synvar{k}_2}{\synvar{vs}}}}
                                    {\synlam{\synvar{vss}}{\synvar{vss}}}}}
   \end{array}
\]

\noindent
CPS-transforming the two local
functions of
Section~\ref{subsec:finding-prefixes-by-accumulating-list-constructors}
adds another layer of continuations and restores the syntactic
characterization of all calls being tail calls and all sub-computations
being elementary.

\subsection{The CPS hierarchy}
\label{subsec:the-CPS-hierarchy}

If $\synvar{k}_2$ were
used non-tail recursively in a variant of the examples of
Section~\ref{subsec:finding-prefixes-in-continuation-passing-style},
we could
CPS-transform the definitions one more time, adding one more layer of
continuations and restoring the syntactic characterization of all
calls being tail calls and all sub-computations being elementary.  We
could also map this definition back to direct style, eliminating
$\synvar{k}_2$ but accessing it with shift.  If the result were mapped
back to direct style one more time, $\synvar{k}_2$ would then be
accessed with a new control operator, shift$_2$, and $\synvar{k}_1$
would be accessed with shift (or more precisely with shift$_1$).

All in all,
successive CPS-transformations induce a CPS
hierarchy~\cite{Danvy-Filinski:LFP90,Danvy-Yang:ESOP99}, and
abstracting control up to each successive layer is achieved with
successive pairs of control operators shift and reset---reset to
initialize the continuation up to a level, and shift to capture a
delimited continuation up to this level.  Each pair of control operators
is indexed by the corresponding level in the hierarchy.  Applying a
captured continuation packages all the current layers on the next layer
and restores the captured layers.  When a captured continuation
completes, the packaged layers are put back into place and the
computation proceeds.  (This informal description is made precise in
Section~\ref{sec:derivation}.)

\subsection{A note about typing}
\label{subsec:a-note-about-typing}

The type of $\synvar{find\_all\_prefixes\_c}_1$, in
Section~\ref{subsec:finding-prefixes-by-accumulating-list-constructors},
is
\[
 \typfun{\typprod{\typfunp{\alpha}{\typbool}}
                 {\typlist{\alpha}}}
        {\typlist{\typlist{\alpha}}}
\]
and the type of its local function $\synvar{visit}$ is
\[
 \typfun{\typprod{\typlist{\alpha}}
                 {\typfunp{\typlist{\alpha}}{\typlist{\alpha}}}}
        {\typlist{\typlist{\alpha}}}.
\]

\noindent
In this example, the co-domain of the continuation is not the same as
the co-domain of $\synvar{visit}$.

Thus 
$\synvar{find\_all\_prefixes\_c}_0$ provides a simple and meaningful example
where Filinski's typing of shift~\cite{Filinski:POPL94}
does not fit, since it must be used at type
\[
 \typfun{\typfunp{\typfunp{\beta}
                          {\synvar{ans}}}
                 {\synvar{ans}}}
        {\beta}
\]
for a given type $\synvar{ans}$, \ie, the answer type of the
continuation and the type of the computation must be the same.  In
other words, control effects are not allowed to change the types of
the contexts. 
Due to a similar restriction on the type of shift, the example does
not fit either in Murthy's pseudo-classical type system for the CPS
hierarchy~\cite{Murthy:CW92} and in Wadler's most general monadic type
system~\cite[Section~3.4]{Wadler:LaSC94-one}.
It however fits in Danvy and Filinski's original type
system~\cite{Danvy-Filinski:DIKU89} which Ariola, Herbelin, and Sabry
have recently embedded in classical subtractive
logic~\cite{Ariola-al:ICFP04}.

\subsection{Related work}
\label{subsection:related-work}

The example considered in this section builds on the simpler function
that unconditionally lists the successive prefixes of a given list.
This simpler function is a traditional example of delimited
continuations~\cite{Danvy:LP89-with-publisher,Sitaram:PhD}:
\begin{itemize}

\item
In the Lisp Pointers~\cite{Danvy:LP89-with-publisher}, Danvy presents three versions of
this function: a typed
continuation-passing version (corresponding to
Section~\ref{subsec:finding-prefixes-by-accumulating-list-constructors}),
one with delimited control (corresponding to
Section~\ref{subsec:finding-prefixes-in-direct-style}), and one in
assembly language.

\item
In his PhD thesis~\cite[Section~6.3]{Sitaram:PhD}, Sitaram presents
two versions of this function: one with an accumulator (corresponding
to Section~\ref{subsec:finding-prefixes-by-accumulating-lists}) and
one with delimited control (corresponding to
Section~\ref{subsec:finding-prefixes-in-direct-style}).

\end{itemize}

\noindent
In
Section~\ref{subsec:finding-prefixes-by-accumulating-list-constructors},
we have shown that the continuation-passing version mediates the version
with an accumulator and the version with delimited control since
defunctionalizing the continu\-ation-passing version yields one and mapping
it back to direct style yields the other.

\subsection{Summary and conclusion}

We have illustrated delimited continuations with the classic example of
finding list prefixes, using CPS as a guideline.  Direct-style programs
using shift and reset can be CPS-transformed into continuation-passing
programs where 
some calls may not be tail calls and some sub-computations may not be
elementary.  One more CPS transformation establishes this syntactic
property with a second layer of continuations.  Further CPS
transformations provide the extra layers of continuation that are
characteristic of the CPS hierarchy.

In the next section, we specify the $\lambda$-calculus extended with
shift and reset.

\begin{figure}[t!]
\hrule
\vspace{2mm}
\begin{itemize}

\itemsp{
Terms:
$
\begin{array}[t]{r@{\ }c@{\ }l}
\typeterm
\ni
\term 
& ::= & 
\synnum{m} 
\Mid
\var
\Mid 
\lam{\var}{\term}
\Mid 
\synapp{\termzero}{\termone}
\Mid
\synsucc{\term}
\Mid
\synreset{\term}
\Mid
\synshift{\vark}{\term}
\end{array}
$}

\itemsp{ 
Values:
$
\begin{array}[t]{r@{\ }c@{\ }l}
\typeval
\ni
\val
& ::= & 
\num{m}
\Mid
f
\end{array}
$}

\item
Answers, 
meta-continuations,
continuations 
and functions: 

\vspace{-2mm}

\[
\begin{array}[t]{rcl}
\ans
& = &
\res
\\
\cnttwo\inp\typecont_2
& = &
\res\rightarrowp\ans
\\
\cntone\inp\typecont_1
& = &
\res\timesp\typecont_2\rightarrowp\ans
\\
f\inp \f
& = &
\res\timesp\typecont_1\timesp\typecont_2\rightarrowp\ans
\end{array}
\]

\item
Initial continuation and meta-continuation:
$
\begin{array}[t]{@{}rcl@{}}
\theta_1
& = &
\lamabs{\pr{\val}{\cnttwo}}
       {\applic{\cnttwo}{\val}}
\\
\theta_2
& = &
\lamabs{\val}
       {\val}
\end{array}
$

\itemsp{ 
Environments:
$
\begin{array}[t]{r@{\ }c@{\ }l}
\typeenv
\ni
\env
& ::= & 
\mtenv
\Mid
\extend{\env}{\var}{\val}
\end{array}
$}

\itemsp{
Evaluation function:
$
\oftype{\eval}{\typeterm\timesp\typeenv\timesp\typecont_1\timesp\typecont_2\rightarrowp\ans}
$
\vspace{2mm}
\[
\begin{array}{@{}rcl@{}}

\namedfour{\eval}
          {\synnum{m}}
          {\env}
          {\cntone}
          {\cnttwo}

& = &

\applic{\cntone}
       {\pr{\num{m}}{\cnttwo}}

\vspace{0.1cm}
\\

\namedfour{\eval}
          {\var}
          {\env}
          {\cntone}
          {\cnttwo}

& = &

\applic{\cntone}
       {\pr{\lookupenv{\var}{\env}}
           {\cnttwo}
         }

\vspace{0.1cm}
\\

\namedfour{\eval}
          {\lam{\var}{\term}}
          {\env}
          {\cntone}
          {\cnttwo}

& = &

\applic{\cntone}
       {\pr{\lamabs{\trip{\val}
                             {\cntone'}
                             {\cnttwo'}
                           }
                     {\namedfour{\eval}
                                {\term}
                                {\extend{\env}{\var}{\val}}
                                {\cntone'}
                                {\cnttwo'}
                              }
                            }
             {\cnttwo}
           }
         
\vspace{0.1cm}
\\

\namedfour{\eval}
          {\synapp{\term_0}{\term_1}}
          {\env}
          {\cntone}
          {\cnttwo}

& = &

\namedfour{\eval}
          {\term_0}
          {\env}
          {\lamabs{\pr{f}{\cnttwo'}}
                  {\namedfour{\eval}
                             {\term_1}
                             {\env}
                             {\lamabs{\pr{\val}{\cnttwo''}}
                                     {\applic{f}{\trip{\val}
                                                      {\cntone}
                                                      {\cnttwo''}}}}
                             {\cnttwo'}}}
          {\cnttwo}
        
\vspace{0.1cm}
\\

\namedfour{\eval}
          {\synsucc{\term}}
          {\env}
          {\cntone}
          {\cnttwo}

& = &

\namedfour{\eval}
          {\term}
          {\env}
          {\lamabs{\pr{\num{m}}{\cnttwo'}}
                  {\applic{\cntone}
                          {\pr{\add{\num{m}}{1}}{\cnttwo'}}
                        }
                }
          {\cnttwo}

\vspace{0.1cm}
\\

\namedfour{\eval}
          {\synreset{\term}}
          {\env}
          {\cntone}
          {\cnttwo}

& = &

\namedfour{\eval}
          {\term}
          {\env}
          {\theta_1}
          {\lamabs{\val}
                  {\applic{\cntone}{\pr{\val}{\cnttwo}}}
                }

\vspace{0.1cm}
\\

\namedfour{\eval}
          {\synshift{\ka}{\term}}
          {\env}
          {\cntone}
          {\cnttwo}

& = &

\namedfour{\eval}
          {\term}
          {\extend{\env}
                  {\ka}
                  {\captcont}}
          {\theta_1}
          {\cnttwo}

\\
&& \text{where}\;
\captcont
= %
\lamabs{\trip{\val}{\cntone'}{\cnttwo'}}
       {\applic{\cntone}{\pr{\val}{\lamabs{\val'}
                                          {\applic{\cntone'}{\pr{\val'}{\cnttwo'}}}}}}

\end{array}
\]}

\item
Main function:
$
\oftype{\main}{\typeterm\rightarrowp\typeval}
$

\[
\begin{array}{@{\hspace{-3.1cm}}rcl@{\hspace{2.5cm}}}

\namedone{\main}{\term}
& = &
\namedfour{\eval}
          {\term}
          {\mtenv}
          {\theta_1}
          {\theta_2}

\end{array}
\]

\end{itemize}
\caption{An environment-based
evaluator for the first
  level of the CPS hierarchy}
\label{fig:def-interpreter-level-one}
\vspace{2mm}
\hrule
\end{figure}

\section{From evaluator to reduction semantics for delimited continuations}
\label{sec:derivation}

\noindent
We derive a reduction semantics for
the call-by-value $\lambda$-calculus extended with shift and reset,
using the method demonstrated in
Section~\ref{sec:arithmetic-expressions}.  First, we transform an
evaluator into an environment-based abstract machine.  Then we
eliminate the environment from this abstract machine, making it
substitution-based.  Finally, we read all the components of a
reduction semantics off the substitution-based abstract machine.

\myindent
Terms
consist of integer literals, variables, $\lambda$-abstractions, function
applications, applications of the successor function, reset expressions,
and shift expressions:
\begin{align*}
\term & ::= \synnum{m} 
\Mid
\var
\Mid 
\lam{\var}{\term}
\Mid 
\synapp{\termzero}{\termone}
\Mid
\synsucc{\term}
\Mid
\synreset{\term}
\Mid
\synshift{\vark}{\term}
\end{align*}
\noindent
Programs are closed terms.

This source language is a subset of the
language used in the examples of
Section~\ref{sec:programming-with-shift-reset}.  Adding
the remaining constructs
is a straightforward
exercise and does not contribute to
our point here.

\subsection{An environment-based
evaluator}
\label{subsec:def-interpreter-env}

\noindent
Figure~\ref{fig:def-interpreter-level-one} displays an
evaluator for the language of the first level of the CPS hierarchy.
This evaluation function represents the original call-by-value semantics
of the $\lambda$-calculus with shift and
reset~\cite{Danvy-Filinski:LFP90}, augmented with integer literals and
applications of the successor function. It is defined by structural
induction over the syntax of terms, and it makes use of an environment
$\env$, a continuation $\cntone$, and a meta-continuation $\cnttwo$.

\myindent
The evaluation of a
terminating program that does not get stuck
(\ie, a program where no ill-formed applications occur in the course of
evaluation) yields either an integer, a function
representing a $\lambda$-abstraction, or a captured continuation.  Both
$\main$ and $\eval$ are partial functions to account for non-terminating
or stuck programs.
The environment stores previously
computed values of the free variables of the term under evaluation.

\myindent The meta-continuation intervenes to interpret reset
expressions and to apply captured continuations.  Otherwise, it is
passively threaded through the evaluation of literals, variables,
$\lambda$-abstractions, function applications, and applications of the
successor function.  (If it were not for shift and reset, and if
$\evalar$
were curried, 
$\cnttwo$ could be eta-reduced and the
evaluator would be in ordinary continuation-passing style.)

\myindent
The reset control operator is used to delimit control.  A reset
expression
\synreset{\term} is interpreted by
evaluating
$\term$ with the initial continuation and a meta-continuation on which
the current continuation has been ``pushed.''  (Indeed, and as will be
shown in Section~\ref{subsec:abstract-machine-env}, defunctionalizing
the meta-continuation yields the data type of a
stack~\cite{Danvy-Nielsen:PPDP01}.)

\myindent
The shift control operator is used to abstract (delimited) control.
A shift expression $\synshift{k}{t}$ is interpreted by capturing the
current continuation, binding it to $k$, and evaluating $t$ in an
environment extended with $k$ and with a continuation reset to the
initial continuation.
Applying a captured continuation is achieved by ``pushing'' the
current continuation on the meta-continuation and applying the
captured continuation to the new meta-continuation.
Resuming a continuation is achieved by reactivating the ``pushed''
continuation with the corresponding meta-continuation.

\subsection{An environment-based abstract machine}
\label{subsec:abstract-machine-env}

\noindent
The
evaluator displayed in Figure~\ref{fig:def-interpreter-level-one} is
already in continuation-passing style.  Therefore, we only need to
defunctionalize its expressible
values and its continuations to obtain an abstract machine.  This
abstract machine is displayed in
Figure~\ref{fig:abstract-machine-env}.

The abstract machine consists of three sets of transitions:
$\mathit{eval}$ for interpreting terms, $\mathit{cont}_1$ for
interpreting the defunctionalized continuations (\ie, the evaluation
contexts),\footnote{The grammar of evaluation contexts in
  Figure~\ref{fig:abstract-machine-env} is isomorphic to the grammar of
  evaluation contexts in the standard
  inside-out notation:
  \[
  \begin{array}{lcl}
    \cont & ::= & \mtcontstd \Mid \argcontstd{\cont}{(\term,\env)} \Mid
    \succcontstd{\cont} \Mid \appcontstd{\val}{\cont}
  \end{array}
  \]
} and $\mathit{cont}_2$ for interpreting the defunctionalized
meta-continuations (\ie, the meta-contexts).\footnote{To build on
  Peyton Jones's terminology~\cite{Marlow-al:ICFP04}, this abstract
  machine is therefore in `eval/apply/meta-apply' form.} The set of possible
values includes integers, closures and captured contexts. In the
original
evaluator, the latter two were represented as higher-order functions, but
defunctionalizing expressible values of the
evaluator has led them to be distinguished.

This eval/apply/meta-apply abstract machine is an extension of the CEK
machine~\cite{Felleisen-Friedman:FDPC3}, which is an eval/apply machine, with the meta-context $\metacont$
and its two transitions, and the two transitions for shift and reset.
$\metacont$ intervenes to process reset expressions and to apply captured
continuations.  Otherwise, it is passively threaded through the
processing of literals, variables, $\lambda$-abstractions, function
applications, and applications of the successor function.  (If it were
not for shift and reset, $\metacont$ and its transitions could be omitted
and the abstract machine would reduce to the CEK machine.)

Given an environment $e$, a context $\cont$, and a meta-context
$\metacont$, a reset expression $\synreset{t}$ is processed by
evaluating $t$ with the same environment $e$, the empty context
$\bullet$, and a meta-context where $\cont$ has been pushed on
$\metacont$.

Given an environment $e$, a context $\cont$, and a meta-context
$\metacont$, a shift expression $\synshift{k}{t}$ is processed by
evaluating $t$ with an extension of $e$ where $k$ denotes $\cont$, the
empty context $\mtcont$, and a meta-context $\metacont$.  Applying a
captured context $\cont'$ is achieved by pushing the current context
$\cont$ on the current meta-context $\metacont$ and continuing with
$\cont'$.  Resuming a context $\cont$ is achieved by popping it off
the meta-context $\conspair{\metacont}{\cont}$ and continuing with
$\cont$.

\begin{figure}[t!]
\hrule
\vspace{2mm}
\begin{itemize}
\itemsp{
Terms:
$
\begin{array}{r@{\ }c@{\ }l}
\term & ::= \synnum{m} 
\Mid
\var
\Mid 
\lam{\var}{\term}
\Mid 
\synapp{\termzero}{\termone}
\Mid
\synsucc{\term}
\Mid
\synreset{\term}
\Mid
\synshift{\vark}{\term}
\end{array}
$}

\itemsp{ 
Values (integers, closures, and captured continuations):
$
\begin{array}{r@{\ }c@{\ }l}
\val
& ::= &
\num{m}
\Mid
\closure{\var}{\term}{\env}
\Mid
\contclosure{\cont}
\end{array}
$}

\itemsp{ 
Environments:
$
\begin{array}{r@{\ }c@{\ }l}
\env
& ::= &
\mtenv
\Mid
\extend{\env}{\var}{\val}
\end{array}
$}

\itemsp{ 
Evaluation contexts: %
$
\begin{array}[t]{r@{\ }c@{\ }l}
\cont
& ::= &
\mtcont
\Mid
\argcont{\cont}{(\term,\env)}
\Mid
\succcont{\cont}
\Mid
\appcont{\val}{\cont}
\end{array}
$}

\itemsp{
Meta-contexts:
$
\begin{array}[t]{r@{\ }c@{\ }l}
\metacont
& ::= &
\mtmetacont
\Mid
\cons{\metacont}{\cont}
\end{array}
$}

\item
Initial transition, transition rules, and final transition:
\[
\hspace{-0.5cm}
\begin{machinetable}

\vspace{3mm}
\spacearound
\transition{\term}
           {\namedquadruple{\evalconfname}
                           {\term}
                           {\mtenv}
                           {\mtcont}
                           {\mtmetacont}
                         }

\spacearound
\transition{\namedquadruple{\evalconfname}
                           {\synnum{m}}
                           {\env}
                           {\cont}
                           {\metacont}
                         }
           {\namedtriple{\contoneconfname}
                        {\cont}
                        {\num{m}}
                        {\metacont}
                      }

\spacearound
\transition{\namedquadruple{\evalconfname}
                           {\var}
                           {\env}
                           {\cont}
                           {\metacont}
                         }
           {\namedtriple{\contoneconfname}
                        {\cont}
                        {\env\,(\var)}
                        {\metacont}
                      }

\spacearound
\transition{\namedquadruple{\evalconfname}
                           {\lam{\var}
                                   {\term}
                                 }
                           {\env}
                           {\cont}
                           {\metacont}
                         }
           {\namedtriple{\contoneconfname}
                        {\cont}
                        {\closure{\var}
                                 {\term}
                                 {\env}
                               }
                        {\metacont}
                      }

\spacearound
\transition{\namedquadruple{\evalconfname}
                           {\termzero\,\termone}
                           {\env}
                           {\cont}
                           {\metacont}
                         }
           {\namedquadruple{\evalconfname}
                           {\termzero}
                           {\env}
                           {\argcont{\cont}
                                    {(\termone,\env)}
                           }
                           {\metacont}
                         } 
 
\spacearound
\transition{\namedquadruple{\evalconfname}
                           {\synsucc{\term}}
                           {\env}
                           {\cont}
                           {\metacont}
                         }  
           {\namedquadruple{\evalconfname}
                           {\term}
                           {\env}
                           {\succcont{\cont}}
                           {\metacont}
                         }

\spacearound
\transition{\namedquadruple{\evalconfname}
                           {\synreset{\term}}
                           {\env}
                           {\cont}
                           {\metacont}
                         }
           {\namedquadruple{\evalconfname}
                           {\term}
                           {\env}
                           {\mtcont}
                           {\conspair{\metacont}
                                 {\cont}
                               }
                         }

\vspace{3mm}
\spacearound
\transition{\namedquadruple{\evalconfname}
                           {\synshift{\vark}{\term}}
                           {\env}
                           {\cont}
                           {\metacont}
                         }
           {\namedquadruple{\evalconfname}
                           {\term}
                           {\extend{\env}
                                   {\vark}
                                   {\cont}}
                           {\mtcont}
                           {\metacont}
                         }

\spacearound
\transition{\namedtriple{\contoneconfname}
                        {\mtcont}
                        {\val}
                        {\metacont}
                      }
           {\namedpair{\conttwoconfname}
                      {\metacont}
                      {\val}
                    }

\spacearound
\transition{\namedtriple{\contoneconfname}
                        {\argcont{\cont}
                                 {(\term,\env)}
                               }
                        {\val}
                        {\metacont}
                      }
           {\namedquadruple{\evalconfname}
                           {\term}
                           {\env}
                           {\appcont{\val}
                                    {\cont}
                                  }
                           {\metacont}
                         }

\spacearound
\hspace{0.2cm}
\transition{\namedtriple{\contoneconfname}
                        {\succcont{\cont}}
                        {\num{m}}
                        {\metacont}
                      }
           {\namedtriple{\contoneconfname}
                        {\cont}
                        {\num{m+1}}
                        {\metacont}
                      }

\spacearound
\transition{\namedtriple{\contoneconfname}
                        {\appcont{\closure{\var}
                                          {\term}
                                          {\env}
                                        }
                                 {\cont}
                               }
                        {\val}
                        {\metacont}
                      }
           {\namedquadruple{\evalconfname}
                           {\term}
                           {\extend{\env}
                                   {\var}
                                   {\val}
                                 }
                           {\cont}
                           {\metacont}
                         }
\vspace{3mm}

\spacearound
\transition{\namedtriple{\contoneconfname}
                        {\appcont{\contclosure{\conttwo}}
                                 {\cont}
                               }
                        {\val}
                        {\metacont}
                      }
           {\namedtriple{\contoneconfname}
                        {\conttwo}
                        {\val}
                        {\conspair{\metacont}
                              {\cont}
                            }
                      }

\vspace{3mm}

\spacearound
\transition{\namedpair{\conttwoconfname}
                      {\conspair{\metacont}
                                {\cont}
                              }
                      {\val}
                    }
           {\namedtriple{\contoneconfname}
                        {\cont}
                        {\val}
                        {\metacont}
                    }

\spacearound
\transition{\namedpair{\conttwoconfname}
                      {\mtmetacont}
                      {\val}
                    }
           {\val}
         
\end{machinetable}
\]
\end{itemize}
\caption{An environment-based abstract machine for the first
  level of the CPS hierarchy}
\label{fig:abstract-machine-env}
\vspace{2mm}
\hrule
\end{figure}

\begin{figure}[t!]
\hrule
\vspace{2mm}
\begin{itemize}

\itemsp{
Terms and values:
$
\begin{array}[t]{r@{\ }c@{\ }l}
\term
& ::= &
\val
\Mid
\var
\Mid 
\synapp{\termzero}{\termone}
\Mid
\synsucc{\term}
\Mid
\synreset{\term}
\Mid
\synshift{\vark}{\term}
\\[1mm]
\val
& ::= &
\synnum{m}
\Mid
\lam{\var}{\term}
\Mid
\cont 
\end{array}
$}

\itemsp{
Evaluation contexts: %
$
\begin{array}[t]{r@{\ }c@{\ }l}
\cont
& ::= &
\mtcont
\Mid
\argcont{\cont}{\term}
\Mid
\succcont{\cont}
\Mid
\appcont{\val}{\cont}
\hspace{1cm}
\end{array}
$}

\itemsp{
Meta-contexts:
$
\begin{array}[t]{r@{\ }c@{\ }l}
\metacont
& ::= &
\mtmetacont
\Mid
\cons{\metacont}{\cont}
\end{array}
$}

\item
Initial transition, transition rules, and final transition:
\[
\begin{machinetable}

\vspace{3mm}
\spacearound
\transition{\term}
           {\namedtriple{\evalconfname}
                        {\term}
                        {\mtcont}
                        {\mtmetacont}
                      }

\spacearound
\transition{\namedtriple{\evalconfname}
                        {\synnum{m}}
                        {\cont}
                        {\metacont}
                      }
           {\namedtriple{\contoneconfname}
                        {\cont}
                        {\synnum{m}}
                        {\metacont}
                      }

\spacearound
\transition{\namedtriple{\evalconfname}
                           {\lam{\var}
                                {\term}
                                 }
                           {\cont}
                           {\metacont}
                         }
           {\namedtriple{\contoneconfname}
                        {\cont}
                        {\lam{\var}
                                {\term}
                              }
                        {\metacont}
                      }
\spacearound
\transition{\namedtriple{\evalconfname}
                        {\conttwo}
                        {\cont}
                        {\metacont}
                      }
           {\namedtriple{\contoneconfname}
                        {\cont}
                        {\conttwo}
                        {\metacont}
                      }

\spacearound
\transition{\namedtriple{\evalconfname}
                           {\synapp{\termzero}
                                   {\termone}
                                 }
                           {\cont}
                           {\metacont}
                         }
           {\namedtriple{\evalconfname}
                        {\termzero}
                        {\argcont{\cont}
                                 {\termone}
                               }
                        {\metacont}
                      } 
 
\spacearound
\transition{\namedtriple{\evalconfname}
                        {\synsucc{\term}}
                        {\cont}
                        {\metacont}
                      }  
           {\namedtriple{\evalconfname}
                        {\term}
                        {\succcont{\cont}}
                        {\metacont}
                      }

\spacearound
\transition{\namedtriple{\evalconfname}
                        {\synreset{\term}}
                        {\cont}
                        {\metacont}
                      }
           {\namedtriple{\evalconfname}
                        {\term}
                        {\mtcont}
                        {\conspair{\metacont}
                                  {\cont}
                            }
                      }
\vspace{3mm}

\spacearound
\transition{\namedtriple{\evalconfname}
                        {\synshift{\vark}{\term}}
                        {\cont}
                        {\metacont}
                      }
           {\namedtriple{\evalconfname}
                        {\subst{\term}
                               {\vark}
                               {\cont}}
                        {\mtcont}
                        {\metacont}
                      }

\spacearound
\transition{\namedtriple{\contoneconfname}
                        {\mtcont}
                        {\val}
                        {\metacont}
                      }
           {\namedpair{\conttwoconfname}
                      {\metacont}
                      {\val}
                    }

\spacearound
\transition{\namedtriple{\contoneconfname}
                        {\argcont{\cont}
                                 {\term}
                               }
                        {\val}
                        {\metacont}
                      }
           {\namedtriple{\evalconfname}
                        {\term}
                        {\appcont{\val}
                                 {\cont}
                               }
                        {\metacont}
                      }

\spacearound
\hspace{0.2cm}
\transition{\namedtriple{\contoneconfname}
                        {\succcont{\cont}}
                        {\synnum{m}}
                        {\metacont}
                      }
           {\namedtriple{\contoneconfname}
                        {\cont}
                        {\synnum{m+1}}
                        {\metacont}
                      }

\spacearound
\transition{\namedtriple{\contoneconfname}
                        {\appcont{\lam{\var}
                                 {\term}
                                       }
                                 {\cont}
                               }
                        {\val}
                        {\metacont}
                      }
           {\namedtriple{\evalconfname}
                        {\subst{\term}{\var}{\val}}
                        {\cont}
                        {\metacont}
                      }

\vspace{3mm}
\spacearound
\transition{\namedtriple{\contoneconfname}
                        {\appcont{\conttwo}
                                 {\cont}
                               }
                        {\val}
                        {\metacont}
                      }
           {\namedtriple{\contoneconfname}
                        {\conttwo}
                        {\val}
                        {\conspair{\metacont}
                                  {\cont}
                            }
                      }

\vspace{3mm}

\spacearound
\transition{\namedpair{\conttwoconfname} 
                      {\conspair{\metacont}
                                {\cont}
                          }
                      {\val}
                    }
           {\namedtriple{\contoneconfname}
                        {\cont}
                        {\val}
                        {\metacont}
                    }

\spacearound
\transition{\namedpair{\conttwoconfname}
                      {\mtmetacont}
                      {\val}
                    }
           {\val}
         
\end{machinetable}
\]
\end{itemize}
\caption{A substitution-based abstract machine for the first
  level of the CPS hierarchy}
\label{fig:abstract-machine-subst}
\vspace{2mm}
\hrule
\end{figure}

The correctness of the abstract machine with respect to the
evaluator
is a consequence of the correctness of
defunctionalization.  In order to express it formally, we define a
partial function $\rawevalenv$ mapping a term $\term$ to a value
$\val$ whenever the environment-based machine, started with
$\term$, stops
with
$\val$. The following theorem states this correctness by relating
observable results:

\begin{theorem}
\label{thm:simulation-level-1-eval-env-am}{\em
  For any
  program $\term$ and any integer value $\num{m}$,
  $\namedone{\main}{\term} = \num{m}$ if and only if $\evalenv{\term}
  = \num{m}$.}
\end{theorem}

\proof
The theorem follows directly from the correctness of
defunctionalization~\cite{Banerjee-al:TACS01,Nielsen:RS-00-47}.
\qed

The environment-based abstract machine can serve both as a foundation
for implementing functional languages with control operators for
delimited continuations and as a stepping stone in theoretical studies
of shift and reset. In the rest of this section, we use it to
construct
a reduction semantics of shift and reset.

\subsection{A substitution-based abstract machine}
\label{subsec:abstract-machine-subst}

The environment-based abstract machine of
Figure~\ref{fig:abstract-machine-env}, on which we want to base our
development, makes a distinction between terms and values.  Since a
reduction semantics is specified by purely syntactic operations (it
gives meaning to terms by specifying their rewriting strategy and an
appropriate notion of reduction, and is indeed also referred to as
`syntactic theory'), we need to embed the domain of values back into
the syntax.  To this end we transform the environment-based abstract
machine into the substitution-based abstract machine displayed in
Figure~\ref{fig:abstract-machine-subst}.  The transformation is
standard, except that we also need to embed evaluation contexts in the
syntax; hence the substitution-based
machine operates on terms where ``quoted'' (in the sense of Lisp)
contexts can occur. (If it were not for shift and reset, $\metacont$
and its transitions could be omitted and the abstract machine would
reduce to the CK machine~\cite{Felleisen-Friedman:FDPC3}.)

We write $\subst{\term}{\var}{\val}$ to denote the result of the usual
capture-avoiding substitution of the value $\val$ for $\var$ in
$\term$.

Formally, the relationship between the two machines is expressed with
the following simulation theorem, where evaluation with the
substitution-based abstract machine is captured by the partial
function $\rawevalsub$, defined analogously to $\rawevalenv$.

\begin{theorem}
\label{thm:simulation-level-1-env-subst-am}{\em
For any program $\term$, either both $\evalsub{\term}$ and
$\evalenv{\term}$ are undefined, or there exist values $\val,\val'$
such that $\evalsub{\term}=\val$, $\evalenv{\term}=\val'$ and
$\real{\val'}=\val$.  The function $\rawreal$
relates a semantic value with its syntactic representation and 
is defined as follows:\footnote{$\rawreal$ is a
  generalization of Plotkin's function Real~\cite{Plotkin:TCS75}.}
\[
\begin{array}{@{}rcl@{}}
  \real{\num{m}} 
  &=&
  \synnum{m}\\[1mm]
  \real{\closure{\var}{\term}{\env}} 
  &=&
  \synlam{\var}{\subst{\subst{\term}{\var_1}{\real{\lookupenv{\var_1}{\env}}}\ldots}{\var_n}{\real{\lookupenv{\var_n}{\env}}}},\\
  &&\text{where }
    \fv{\lamabs{\var}{\term}}=\{\var_1,\ldots,\var_n\}\\[1mm]
  \real{\mtcont} 
  &=&
  \mtcont\\[1mm]
  \real{\argcont{\cont}{(\term,\env)}}
  &=&
  \argcont{\real{\cont}}{\subst{\subst{\term}{\var_1}{\real{\lookupenv{\var_1}{\env}}}\ldots}{\var_n}{\real{\lookupenv{\var_n}{\env}}}},\\
  &&\text{where }
  \fv{\term} = \{\var_1,\ldots,\var_n\} \\[1mm]
  \real{\appcont{\val}{\cont}} 
  &=&
  \appcont{\real{\val}}{\real{\cont}} \\[1mm]
  \real{\succcont{\cont}} 
  &=&
  \succcont{\real{\cont}}
\end{array}
\]}
\end{theorem}

\proof
We extend the translation function $\rawreal$ to meta-contexts and
configurations, in the expected way, \eg,
\[
\begin{array}{lcl}
\real{\namedquadruple{\evalconfname}
                     {\term}
                     {\env}
                     {\cont}
                     {\metacont}} 
&=&
\namedtriple{\evalconfname}
            {\subst{\subst{\term}{\var_1}{\real{\lookupenv{\var_1}{\env}}}\ldots}{\var_n}{\real{\lookupenv{\var_n}{\env}}}}
            {\real{\cont}}
            {\real{\metacont}}
\\
&& \text{where } \fv{\term}=\{\var_1,\ldots,\var_n\}
\end{array}
\]
Then it is straightforward to show that the two abstract machines
operate in lock step with respect to the translation. Hence, for any
program $\term$, both machines diverge or they both stop (after the
same number of transitions) with the values $\val$ and $\real{\val}$,
respectively.
\qed

\myindent
We now proceed to analyze the transitions of the machine displayed in
Figure~\ref{fig:abstract-machine-subst}.
We can think of a configuration
$\namedtriple{eval}{\term}{\cont}{\metacont}$ as the following
decomposition of the initial term into a meta-context \metacont, a
context \cont, and an intermediate term \term:
$$\decomptwo{\metacont}{\cont}{\term}$$
\noindent
where\del\ separates the context and the meta-context.
Each transition performs either a reduction, or a
decomposition in search of the next redex. Let us recall that a
decomposition is performed when both sides of a transition are partitions
of the same term; in that case, depending on the structure of the
decomposition $\decomptwo{\metacont}{\cont}{\term}$, a subpart of the
term is chosen to be evaluated next, and the contexts are updated
accordingly. We also observe that $\mathit{eval}$-transitions follow the
structure of $\term$, $\mathit{cont}_1$-transitions follow the structure
of $\cont$ when the term has been reduced to a value, and
$\mathit{cont}_2$-transitions follow the structure of $\metacont$ when a
value in the empty context has been reached.

Next we specify all the components of the reduction semantics based on
the analysis of the abstract machine.

\subsection{A reduction semantics}
\label{subsec:a-reduction-semantics}

\myindent
A reduction semantics provides a reduction relation on expressions by
defining values, evaluation contexts, and
redexes~\cite{Felleisen:PhD,Felleisen-Flatt:LN,Felleisen-Friedman:FDPC3,Xiao-al:HOSC01}.
In the present case,
\begin{itemize}
  
\item the values are already specified in the (substitution-based)
  abstract machine:
\begin{align*}
  \val ::= \synnum{m} \Mid \lam{\var}{\term} \Mid \cont
\end{align*}

\item the evaluation contexts and meta-contexts are already specified in
the abstract machine, as the data-type part of defunctionalized
continuations;
\begin{align*}
  \cont & ::= \mtcont \Mid \argcont{\cont}{\term} \Mid
  \appcont{\val}{\cont} \Mid \succcont{\cont}
  \\
  \metacont & ::= \mtmetacont \Mid \cons{\metacont}{\cont}
\end{align*}

\item we can read the 
redexes off the transitions of the abstract
machine: %
\begin{align*}
  \redex & ::= \synsucc{\synnum{m}}
  \Mid
  \synapp{\lamp{\var}{\term}}{\val}
  \Mid
  \synshift{\vark}{\term}
  \Mid
  \synapp{\conttwo}{\val}
  \Mid
  \synreset{\val}
\end{align*}
\end{itemize}

Based on the distinction between decomposition and reduction, we
single out the following reduction rules from the transitions of the
machine: %

\[
\begin{array}{lrl}
  (\delta)\;\; & 
  \decomptwo{\metacont}{\cont}{\synsucc{\synnum{m}}} &
  \red\;\; \decomptwo{\metacont}{\cont}{\synnum{m+1}}
  \\[1mm]
  (\beta_{\lambda})\;\; &
  \decomptwo{\metacont}{\cont}{\synapp{\lamp{\var}{\term}}{\val}} & \red\;\;
  \decomptwo{\metacont}{\cont}{\subst{\term}{\var}{\val}}
  \\[1mm]
   (\rawshift_{\lambda})\;\; &
  \decomptwo{\metacont}{\cont}{\synshift{\vark}{\term}} & \red\;\;
  \decomptwo{\metacont}{}{\subst{\term}{\vark}{\cont}}  %
  \\[1mm]
  (\beta_{\mathit{ctx}})\;\; &
  \decomptwo{\metacont}{\cont}{\synapp{\conttwo}{\val}} & \red\;\;
  \decomptwo{\cons{\metacont}{\cont}}{\conttwo}{\val}
  \\[1mm]
  (\text{Reset})\;\; &
  \decomptwo{\metacont}{\cont}{\synreset{\val}} & \red\;\;
  \decomptwo{\metacont}{\cont}{\val}
\end{array}
\]
\vspace{1mm}

\noindent
($\beta_{\lambda}$) is the usual call-by-value $\beta$-reduction; we have
renamed it to indicate that the applied term is a $\lambda$-abstraction,
since we can also apply a captured context, as in
($\beta_{\mathit{ctx}}$).
($\rawshift_{\lambda}$) is plausibly symmetric to
($\beta_{\lambda}$) --- it can be seen as an application of the
abstraction $\lam{\vark}{\term}$
to the current context.
Moreover,
($\beta_{\mathit{ctx}}$) can be seen as performing both a reduction and a
decomposition: it is a reduction because an application of a context with
a hole to a value is reduced to the value plugged into the hole; and it
is a decomposition because it changes the meta-context, as if the
application were enclosed in a reset.  Finally,
(Reset) makes it possible to pass the boundary of a context when the term
inside this context has been reduced to a value.

\myindent
The $\beta_{\mathit{ctx}}$-rule and the $\rawshift_{\lambda}$-rule
give a justification for representing a captured context $\cont$ as a
term $\lam{\var}{\synreset{\cont[\var]}}$, as found in other studies
of shift and reset~\cite{Kameyama:CSL04, Kameyama-Hasegawa:ICFP03,
Murthy:CW92}. In
particular, the need for delimiting the captured context is a
consequence of the $\beta_{\mathit{ctx}}$-rule.
\myindent
Finally, we can read the decomposition function off the
transitions of the abstract machine:
\[
\begin{array}{rcl}
\decompose{\term} &=& \dec{\term}{\mtcont}{\mtmetacont}\\
\dec{\termzero\,\termone}{\cont}{\metacont} &=&
\dec{\termzero}{\argcont{\cont}{\termone}}{\metacont}\\
\dec{\synsucc{\term}}{\cont}{\metacont} &=&
\dec{\term}{\succcont{\cont}}{\metacont}\\
\dec{\synreset{\term}}{\cont}{\metacont} &=&
\dec{\term}{\mtcont}{\cons{\metacont}{\cont}}\\
\dec{\val}{\argcont{\cont}{\term}}{\metacont} &=&
\dec{\term}{\appcont{\val}{\cont}}{\metacont}\\[2mm]
\end{array}
\]

\noindent
In the remaining cases either a value or a 
redex has been found:
\[
\begin{array}{rcl}
\dec{\val}{\mtcont}{\mtmetacont} &=&
\decomptwo{\mtmetacont}{}{\val}\\ %
\dec{\val}{\mtcont}{\cons{\metacont}{\cont}} &=&
\decomptwo{\metacont}{\cont}{\synreset{\val}}\\
\dec{\synshift{\vark}{\term}}{\cont}{\metacont} &=&
\decomptwo{\metacont}{\cont}{\synshift{\vark}{\term}}\\
\dec{\val}{\appcont{\lamp{\var}{\term}}{\cont}}{\metacont} &=&
\decomptwo{\metacont}{\cont}{\synapp{(\lam{\var}{\term})}{\val}}\\
\dec{\val}{\appcont{\conttwo}{\cont}}{\metacont} &=&
\decomptwo{\metacont}{\cont}{\synapp{\conttwo}{\val}}\\
\dec{\synnum{m}}{\succcont{\cont}}{\metacont} &=&
\decomptwo{\metacont}{\cont}{\synsucc{\synnum{m}}}
\end{array}
\]

\myindent
An inverse of the $\rawdecompose$ function, traditionally called
$\rawplug$, reconstructs a term from its decomposition:
\[ 
\begin{array}{rcl}
\plug{\decomptwo{\mtmetacont}{}{\term}} %
& = &
\term
\\
\plug{\decomptwo{\conspair{\metacont}{\cont}}{}{\term}} %
& = &
\plug{\decomptwo{\metacont}{\cont}{\synreset{\term}}}
\\
\plug{\decomptwo{\metacont}{(\argcont{\cont}{\term'})}{\term}}
& = &
\plug{\decomptwo{\metacont}{\cont}{\synapp{\term}{\term'}}}
\\
\plug{\decomptwo{\metacont}{(\appcont{\val}{\cont})}{\term}}
& = &
\plug{\decomptwo{\metacont}{\cont}{\synapp{\val}{\term}}}
\\
\plug{\decomptwo{\metacont}{(\succcont{\cont})}{\term}}
& = &
\plug{\decomptwo{\metacont}{\cont}{\synsucc{\term}}}
\end{array}
\]

\myindent
In order to talk about unique decomposition, we need to define the set
of potential redexes (\ie, the disjoint union of actual redexes
and stuck redexes). The grammar of potential redexes reads as follows:
\begin{align*}
  \potredex & ::= \synsucc{\val} 
  \Mid
  \synapp{\val_0}{\val_1} 
  \Mid 
  \synshift{\vark}{\term} 
  \Mid
  \synreset{\val}
\end{align*}

\begin{lemma}[Unique decomposition]
  A program $\term$ is either a value $\val$ or there exist a unique
  context $\cont$, a unique meta-context $\metacont$ and a potential
  redex $\potredex$ such that $\term =
  \plug{\decomptwo{\metacont}{\cont}{\potredex}}$.  In the former case
  $\decompose{\term} = \decomptwo{\mtmetacont}{}{\val}$ and in the
  latter case either $\decompose{\term} =
  \decomptwo{\metacont}{\cont}{\potredex}$ if $\potredex$ is an actual
  redex, or $\decompose{\term}$ is undefined.
\end{lemma}

\proof
  The first part follows by induction on the structure of $\term$.
  The second part follows from the equation
  $\decompose{\plug{\decomptwo{\metacont}{\cont}{\redex}}} =
  \decomptwo{\metacont}{\cont}{\redex}$ which holds for all
  $\metacont$, $\cont$ and $\redex$.
\qed

It is evident that evaluating a program either using the derived
reduction rules or using the substitution-based abstract machine
yields the same result.

\begin{theorem}
{\em For any program $\term$ and any value $\val$, $\evalsub{\term} = \val$
if and only if $\term\redtrans\val$, where $\redtrans$ is the
reflexive, transitive closure of the one-step reduction defined by
the relation $\red$.}
\end{theorem}

\proof
When evaluating with the abstract machine, each contraction is
followed by decomposing the contractum in the current context and
meta-context. When evaluating with the reduction rules, however,
each contraction is followed by plugging the contractum and
decomposing the resulting term.
Therefore, the theorem follows from the equation 
\[\dec{\term}{\cont}{\metacont} = 
\decompose{\plug{\decomptwo{\metacont}{\cont}{\term}}}\]
which holds for any $\metacont$, $\cont$ and $\term$.
\qed

We have verified that using refocusing~\cite{Biernacka-Danvy:RS-05-22,Danvy-Nielsen:RS-04-26},
one can go from
this reduction semantics
to the 
abstract machine of Figure~\ref{fig:abstract-machine-subst}.

\subsection{Beyond CPS}
\label{subsec:beyond-CPS}

Alternatively to using the meta-context to compose delimited
continuations, as in Figure~\ref{fig:abstract-machine-env},
we could compose
them by concatenating their representation~\cite{Felleisen-al:LFP88}.  Such a concatenation
function is defined as follows:
\begin{eqnarray*}
\concatcont{\mtcont}{\cont'}
& = &
\cont'
\\
\concatcont{\argcontp{\cont}{(\term,\env)}}{\cont'}
& = &
\argcont{\concatcont{\cont}{\cont'}}{(\term,\env)}
\\
\concatcont{\succcontp{\cont}}{\cont'}
& = &
\succcont{\concatcont{\cont}{\cont'}}
\\
\concatcont{\appcontp{\val}{\cont}}{\cont'}
& = &
\appcont{\val}{\concatcont{\cont}{\cont'}}
\end{eqnarray*}

\noindent
(The second
clause would read $\concatcont{\argcontp{\cont}{\term}}{\cont'} =
\argcont{\concatcont{\cont}{\cont'}}{\term}$
for the contexts of Figure~\ref{fig:abstract-machine-subst}.)

Then, in Figures~\ref{fig:abstract-machine-env} and
\ref{fig:abstract-machine-subst}, we could replace the transition

\[
\begin{machinetable}

\spacearound
\transition{\namedtriple{\contoneconfname}
                        {\appcont{\contclosure{\conttwo}}
                                 {\cont}
                               }
                        {\val}
                        {\metacont}
                      }
           {\namedtriple{\contoneconfname}
                        {\conttwo}
                        {\val}
                        {\conspair{\metacont}
                              {\cont}
                            }
                      }
\end{machinetable}
\]

\vspace{1mm}

\noindent
by the following one:

\[
\begin{machinetable}

\spacearound
\transition{\namedtriple{\contoneconfname}
                        {\appcont{\contclosure{\conttwo}}
                                 {\cont}
                               }
                        {\val}
                        {\metacont}
                      }
           {\namedtriple{\contoneconfname}
                        {\concatcont{\conttwo}{\cont}}
                        {\val}
                        {\metacont}
                      }
\end{machinetable}
\]

\vspace{1mm}

\noindent
This replacement changes the control effect of shift to that of Felleisen
et al.'s $\mathcal{F}$ operator~\cite{Felleisen:POPL88}.  Furthermore,
the modified abstract machine is in structural correspondence with
Felleisen et al.'s abstract machine for $\mathcal{F}$ and
$\#$~\cite{Felleisen:POPL88,Felleisen-al:LFP88}.

This representation of control (as a list of `stack frames') and this
implementation of composing delimited continuations (by concatenating
these lists) are at the heart of virtually all non-CPS-based accounts of
delimited control.  However, the modified environment-based abstract
machine does not correspond to a defunctionalized continuation-passing
evaluator because it is not in the range of
defunctionalization~\cite{Danvy-Nielsen:PPDP01} since the first-order
representation of functions should have a single point of consumption.
Here, the constructors of contexts are not solely consumed by the
$\contoneconfname$ transitions of the abstract machine as in
Figures~\ref{fig:abstract-machine-env} and
\ref{fig:abstract-machine-subst}, but also by $\rawconcatcont$.  Therefore,
the
abstract machine that uses $\rawconcatcont$ is not in the range of
Reynolds's
defunctionalization and it thus does not immediately
correspond to a higher-order,
continuation-passing evaluator.  In that sense, control operators using
$\rawconcatcont$ go beyond CPS.

Elsewhere~\cite{Biernacki-al:RS-05-16}, we have rephrased the modified
abstract machine to put it in defunctionalized form, and we have
exhibited the corresponding higher-order evaluator and the corresponding
`dynamic' continuation-passing style.  This dynamic CPS is not just plain CPS but
is a form of continuation+state-passing style where the threaded state is
a list of intermediate delimited continuations.  Unexpectedly, it is also
in structural correspondence with the architecture for delimited control
recently proposed by Dybvig, Peyton Jones, and Sabry on other operational
grounds~\cite{Dybvig-al:TR05}.

\subsection{Static vs.~dynamic delimited continuations}
\label{subsec:static-vs-dynamic-del-conts}

Irrespectively of any new dynamic CPS and any new
architecture for delimited control, there seems to be remarkably few
examples that actually illustrate the expressive power of dynamic
delimited continuations.  We have recently presented one, breadth-first
traversal~\cite{Biernacki-al:IPL05}, and we present another one
below.

The two following functions traverse a given list and return another
list.  The recursive call to $\synvar{visit}$ is abstracted into a
delimited continuation, which is applied to the tail of the list:
\[
\begin{array}{rcl}
   \synapp{\synvar{foo}}
          {\synvar{xs}}
   &
   \stackrel{\mathrm{def}}{=}
   &
   \vsynletrectwo{\synapp{\synvar{visit}}
                         {\synnil}}
                 {\synnil}
                 {\synapp{\synvar{visit}}
                         {\synconsp{\synvar{x}}
                                   {\synvar{xs}}}}
                 {\synapp{\synvar{visit}}
                         {\synshiftp{\synvar{k}}
                                    {\syncons{x}
                                             {\synappp{\synvar{k}}
                                                      {\synvar{xs}}}}}}
                 {\synreset{\synapp{\synvar{visit}}{\synvar{xs}}}}
\end{array}
\mbox{\hspace{1cm}}
\begin{array}{rcl}
   \synapp{\synvar{bar}}
          {\synvar{xs}}
   &
   \stackrel{\mathrm{def}}{=}
   &
   \vsynletrectwo{\synapp{\synvar{visit}}
                         {\synnil}}
                 {\synnil}
                 {\synapp{\synvar{visit}}
                         {\synconsp{\synvar{x}}
                                   {\synvar{xs}}}}
                 {\synapp{\synvar{visit}}
                         {\syncontrolp{\synvar{k}}
                                      {\syncons{x}
                                               {\synappp{\synvar{k}}
                                                        {\synvar{xs}}}}}}
                 {\synreset{\synapp{\synvar{visit}}{\synvar{xs}}}}
\end{array}
\]

\newcommand{\myunderline}[1]{$\underline{\mbox{#1}}$}

\noindent
On the left, $\synvar{foo}$ uses $\mathcal{S}$ and on the right,
$\synvar{bar}$ uses $\mathcal{F}$; for the rest, the two definitions are
identical.  Given an input list, $\synvar{foo}$ 
\myunderline{co}p\myunderline{ies}
it and $\synvar{bar}$
\myunderline{reverses}
it.

\newcommand{\conscont}[2]{\ensuremath{\mathssf{CONS}(#1, #2)}}
\newcommand{\conscontp}[2]{(\conscont{#1}{#2})}

To explain this difference and to account for the extended source
language, we need to expand the grammar of evaluation
contexts, \eg, with a production to account for calls to the list
constructor:
\[
\begin{array}[t]{r@{\ }c@{\ }l}
\cont
& ::= &
\mtcont
\Mid
\argcont{\cont}{\term}
\Mid
\succcont{\cont}
\Mid
\appcont{\val}{\cont}
\Mid
\conscont{\val}{\cont}
\Mid
\ldots
\end{array}
\]

\noindent
Similarly, we need to expand the definition of concatenation as follows:
\begin{eqnarray*}
\concatcont{\conscontp{\val}{\cont}}{\cont'}
& = &
\conscont{\val}{\concatcont{\cont}{\cont'}}
\end{eqnarray*}

\newcommand{\sepdecomptwo}[3]{{#1}&\del&{#2}[{#3}]}

Here is a trace of the two computations in the form of the calls to and
returns from $\synvar{visit}$ for the input list
$\syncons{1}{\syncons{2}{\synnil}}$:

\begin{description}

\item[\emph{foo}]
Every time the captured continuation is resumed, its representation is
kept separate from the current context.  The meta-context therefore grows
whereas the captured context solely consists of
$\appcont{\synvar{visit}}{\mtcont}$ throughout (writing $\synvar{visit}$
in the context for simplicity):
\[
  \begin{array}{@{}r@{\hspace{0.5mm}}c@{\hspace{0.5mm}}l@{\hspace{3.3cm}}}
\sepdecomptwo{\metacont}
            {\cont}
            {\synreset{\synapp{\synvar{visit}}{\synconsp{1}{\syncons{2}{\synnil}}}}}
  \\
\sepdecomptwo{\cons{\metacont}{\cont}}
            {}
            {\synapp{\synvar{visit}}{\synconsp{1}{\syncons{2}{\synnil}}}}
  \\
\sepdecomptwo{\cons{\cons{\metacont}{\cont}}{\conscontp{1}{\mtcont}}}
            {}
            {\synapp{\synvar{visit}}{\synconsp{2}{\synnil}}}
  \\
\sepdecomptwo{\cons{\cons{\cons{\metacont}{\cont}}{\conscontp{1}{\mtcont}}}
                  {\conscontp{2}{\mtcont}}}
            {}
            {\synapp{\synvar{visit}}{\synnil}}
  \\
\sepdecomptwo{\cons{\cons{\cons{\metacont}{\cont}}{\conscontp{1}{\mtcont}}}
                  {\conscontp{2}{\mtcont}}}
            {}
            {\synnil}
  \\
\sepdecomptwo{\cons{\cons{\metacont}{\cont}}{\conscontp{1}{\mtcont}}}
            {}
            {\syncons{2}{\synnil}}
  \\
\sepdecomptwo{\cons{\metacont}{\cont}}
            {}
            {\syncons{1}{\syncons{2}{\synnil}}}
  \\
\sepdecomptwo{\metacont}
            {\cont}
            {\syncons{1}{\syncons{2}{\synnil}}}
  \end{array}
\]

\item[\hspace{-1.5mm}\emph{bar}]
Every time the captured continuation is resumed, its representation is
concatenated to the current context.
The meta-context therefore remains the same whereas the
context changes dynamically.  The first captured context is
$\appcont{\synvar{visit}}{\mtcont}$; concatenating it to
$\conscont{1}{\mtcont}$ yields
$\conscont{1}{\appcont{\synvar{visit}}{\mtcont}}$, which is the second
captured context:
\[
  \begin{array}{@{\hspace{3.3cm}}r@{\hspace{0.5mm}}c@{\hspace{0.5mm}}l}
\sepdecomptwo{\metacont}
            {\cont}
            {\synreset{\synapp{\synvar{visit}}{\synconsp{1}{\syncons{2}{\synnil}}}}}
  \\
\sepdecomptwo{\cons{\metacont}{\cont}}
            {}
            {\synapp{\synvar{visit}}{\synconsp{1}{\syncons{2}{\synnil}}}}
  \\
\sepdecomptwo{\cons{\metacont}{\cont}}
            {\conscontp{1}{\mtcont}}
            {\synapp{\synvar{visit}}{\synconsp{2}{\synnil}}}
  \\
\sepdecomptwo{\cons{\metacont}{\cont}}
            {\conscontp{2}{\conscont{1}{\mtcont}}}
            {\synapp{\synvar{visit}}{\synnil}}
  \\
\sepdecomptwo{\cons{\metacont}{\cont}}
            {\conscontp{2}{\conscont{1}{\mtcont}}}
            {\synnil}
  \\
\sepdecomptwo{\cons{\metacont}{\cont}}
            {\conscontp{2}{\mtcont}}
            {\syncons{1}{\synnil}}
  \\
\sepdecomptwo{\cons{\metacont}{\cont}}
            {}
            {\syncons{2}{\syncons{1}{\synnil}}}
  \\
\sepdecomptwo{\metacont}
            {\cont}
            {\syncons{2}{\syncons{1}{\synnil}}}
  \end{array}
\]

\end{description}

\subsection{Summary and conclusion}

We have presented the original
evaluator for the $\lambda$-calculus
with shift and reset; this evaluator uses two layers of continuations.
From this call-by-value
evaluator we have derived two abstract machines, an environment-based
one and a substitution-based one; each of these machines uses two
layers of evaluation contexts.  Based on the substitution-based
machine we have constructed a reduction semantics for the
$\lambda$-calculus with shift and reset;
this reduction semantics, by construction, is
sound with respect to CPS.
Finally, we have pointed out the difference between the static and
dynamic delimited control operators at the level of the abstract
machine and we have presented a simple but meaningful example
illustrating their differing behavior.

\section{From evaluator to reduction semantics for the CPS hierarchy}
\label{sec:up-in-the-hierarchy}

\noindent
We construct a reduction semantics for the call-by-value
$\lambda$-calculus extended with shift$_n$ and reset$_n$.  As in
Section~\ref{sec:derivation}, we go from
an evaluator to an environment-based abstract machine, and from a
substitution-based abstract machine to a reduction semantics.
Because of the regularity of CPS, the results can be generalized from
level $1$ to higher levels
without repeating the actual
construction, based only on the original specification of the
hierarchy~\cite{Danvy-Filinski:LFP90}. In particular, the proofs of
the theorems generalize straightforwardly from level 1.

\begin{figure}[b!!]%
\hrule
\vspace{2mm}
\begin{itemize}

\itemsp{
Terms ($1 \leq i \leq n$):
\hspace{-0.3cm}
$
\begin{array}{r@{\ }c@{\ }l@{}}
\typeterm 
\ni
\term 
& ::= & 
\synnum{m} 
\Mid
\var
\Mid 
\lam{\var}{\term}
\Mid 
\synapp{\termzero}{\termone}
\Mid
\synsucc{\term}
\Mid
\synresetn{i}{\term}
\Mid
\synshiftn{i}{\vark}{\term}
\end{array}
$}

\itemsp{
Values:
$
\begin{array}[t]{r@{\ }c@{\ }l}
\typeval
\ni
\val
& ::= &
\num{m}
\Mid
f
\end{array}
$}

\item
Answers, continuations and functions ($1 \leq i \leq n$):
\[
\begin{array}{rclcl}
& & \ans & = & \res 
\\
\kn{n+1} & \in & \typecont_{n+1} 
& = &
\res\rightarrowp\ans\\
\kn{i} & \in & \typecont_{i} 
& = &
\res\timesp\typecont_{i+1}\timesp\dots\timesp\typecont_{n+1}\rightarrowp\ans

\\
f & \in & \f & = &
\res\timesp\typecont_{1\mbox{\hspace{3.4mm}}}\timesp\dots\timesp\typecont_{n+1}\rightarrowp\ans
\end{array}
\]

\item
Initial continuations ($1 \leq i \leq n$):
\[
\begin{array}[t]{r@{\ }c@{\ }l}
\theta_i 
& \equals &
\lamabs{\quin{\val}{\kn{i+1}}{\kn{i+2}}{\dots}{\kn{n+1}}}
       {\applic{\kn{i+1}}{\quadr{\val}{\kn{i+2}}{\dots}{\kn{n+1}}}}
\\
\theta_{n+1}
& = &
\lamabs{\val}
       {\val}
\end{array}
\]

\itemsp{ 
Environments:
$
\begin{array}[t]{r@{\ }c@{\ }l}
\typeenv
\ni
\env
& ::= &
\mtenv
\Mid
\extend{\env}{\var}{\val}
\end{array}
$}

\item
Evaluation function:
see Figure~\ref{fig:def-interpreter-level-n-ctd}

\end{itemize}
\caption{An environment-based
evaluator for the CPS
  hierarchy at level $n$}
\label{fig:def-interpreter-level-n}
\vspace{2mm}
\hrule
\end{figure}

\begin{sidewaysfigure}
\hrule
\vspace{2mm}
\begin{itemize}
\item
Evaluation function ($1 \leq i \leq n$):
$
\oftype{\evaln}{\typeterm\timesp\typeenv\timesp\typecont_1\timesp\dots\timesp\typecont_{n+1}\rightarrowp\ans}
$
\vspace{2mm}
\[
\begin{array}{@{\hspace{-0.9cm}}r@{\ \ }c@{\ \ }l@{\hspace{-0.9cm}}}
\vspace{0.1cm}
\namedsix{\evaln}
         {\synnum{m}}
         {\env}
         {\kn{1}}
         {\kn{2}}
         {...}%
         {\kn{n+1}}

& = &

\applic{\kn{1}}
       {\quadr{\num{m}}{\kn{2}}{...}{\kn{n+1}}}

\vspace{0.1cm}
\\

\namedsix{\evaln}
         {\var}
         {\env}
         {\kn{1}}
         {\kn{2}}
         {...}%
         {\kn{n+1}}

& = &

\applic{\cntone}
       {\quadr{\lookupenv{\var}{\env}}{\kn{2}}{...}{\kn{n+1}}}

\vspace{0.1cm}
\\

\namedsix{\evaln}
         {\lam{\var}{\term}}
         {\env}
         {\kn{1}}
         {\kn{2}}
         {...}%
         {\kn{n+1}}

& = &

\applic{\kn{1}}
       {\quadr{\lamabs{\quin{\val}{\kn{1}'}{\kn{2}'}{...}{\kn{n+1}'}}
                      {\namedsix{\evaln}
                                {\term}
                                {\extend{\env}{\var}{\val}}
                                {\kn{1}'}
                                {\kn{2}'}
                                {...}%
                                {\kn{n+1}'}}}
             {\kn{2}}
             {...}%
             {\kn{n+1}}}

\vspace{0.1cm}
\\

\namedsix{\evaln}
         {\synapp{\term_0}{\term_1}}
         {\env}
         {\kn{1}}
         {\kn{2}}
         {...}%
         {\kn{n+1}}

& = &

\namedtwolp{\evaln}{\term_0}{\env}
\\
& &
\hspace{0.95cm}
\lamabs{\quadr{f}{\kn{2}'}{...}{\kn{n+1}'}}
       {\namedtwolp{\evaln}{\term_1}{\env}}
\\
& &
\hspace{5.10cm}
\lamabs{\quadr{\val}{\kn{2}''}{...}{\kn{n+1}''}}
       {\applic{f}
               {\quin{\val}{\kn{1}}{\kn{2}''}{...}{\kn{n+1}''}}},
\\
& &
\hspace{5.10cm}
\triprp{\kn{2}'}{...}{\kn{n+1}'},
\\
& &
\hspace{0.95cm}
\triprp{\kn{2}}{...}{\kn{n+1}}

\vspace{0.1cm}
\\

\namedsix{\evaln}
         {\synsucc{\term}}
         {\env}
         {\kn{1}}
         {\kn{2}}
         {...}%
         {\kn{n+1}}

& = &

\namedsix{\evaln}
         {\term}
         {\env}
         {\lamabs{\quadr{\num{m}}{\kn{2}'}{...}{\kn{n+1}'}}
                 {\applic{\kn{1}}
                         {\quadr{\add{\num{m}}{1}}{\kn{2}'}{...}{\kn{n+1}'}}}}
         {\kn{2}}
         {...}%
         {\kn{n+1}}

\vspace{0.1cm}
\\

\namedsix{\evaln}
         {\synresetn{i}{\term}}
         {\env}
         {\kn{1}}
         {\kn{2}}
         {...}%
         {\kn{n+1}}

& = &

\namedeight{\evaln}
           {\term}
           {\env}
           {\theta_1,\,\dots}
           {\theta_i}
           {\lamabs{\quadr{\val}{\kn{i+2}'}{...}{\kn{n+1}'}}
                   {\applic{\kn{1}}
                           {\sev{\val}{\kn{2}}{...}{\kn{i+1}}{\kn{i+2}'}{...}{\kn{n+1}'}}}}
           {\kn{i+2}}
           {...}%
           {\kn{n+1}}

\vspace{0.1cm}
\\

\namedsix{\evaln}
         {\synshiftn{i}{\ka}{\term}}
         {\env}
         {\kn{1}}
         {\kn{2}}
         {...}%
         {\kn{n+1}}

& = &

\namedeight{\evaln}
           {\term}
           {\extend{\env}{\ka}{\captcont_i}}
           {\theta_1}
           {...}%
           {\theta_i}
           {\kn{i+1}}
           {...}%
           {\kn{n+1}}

\end{array}
\]
\[
\text{where}\;
\captcont_i
= %
\lamabs{\quadr{\val}{\kn{1}'}{...}{\kn{n+1}'}}
       {\applic{\kn{1}}
               {\eight{\val}
                      {\kn{2}}
                      {...}%
                      {\kn{i}}
                      {\lamabs{\quadr{\val'}{\kn{i+2}''}{...}{\kn{n+1}''}}
                              {\applic{\kn{1}'}
                                      {\sev{\val'}
                                           {\kn{2}'}
                                           {...}%
                                           {\kn{i+1}'}
                                           {\kn{i+2}''}
                                           {...}%
                                           {\kn{n+1}''}}}}
                      {\kn{i+2}'}
                      {...}%
                      {\kn{n+1}'}}}
\]
\item
Main function: $\oftype{\mainn}{\typeterm\rightarrowp\typeval}$
\[
\begin{array}{rcl}
\namedone{\mainn}{\term}
& = &
\namedsix{\evaln}
         {\term}
         {\mtenv}
         {\theta_1}
         {...}%
         {\theta_n}
         {\theta_{n+1}}
\hspace{5.3cm}
\end{array}
\]
\end{itemize}
\caption{An environment-based
evaluator for the CPS hierarchy
  at level $n$, ctd.}
\label{fig:def-interpreter-level-n-ctd}
\vspace{2mm}
\hrule
\end{sidewaysfigure}

\begin{figure}[b!!]%
\hrule
\vspace{2mm}
\begin{itemize}
\itemsp{
Terms ($1 \leq i \leq n$):
$
\begin{array}[t]{r@{\ }c@{\ }l}
\term 
& ::= & 
\synnum{m} 
\Mid
\var
\Mid 
\lam{\var}{\term}
\Mid 
\synapp{\termzero}{\termone}
\Mid
\synsucc{\term}
\Mid
\synresetn{i}{\term}
\Mid
\synshiftn{i}{\vark}{\term}
\end{array}
$}
\itemsp{ 
Values ($1 \leq i \leq n$):
$
\begin{array}[t]{r@{\ }c@{\ }l}
\val
& ::= &
\num{m}
\Mid
\closure{\var}{\term}{\env}
\Mid
\cn{i}
\end{array}
$}
\item
Evaluation contexts ($2 \leq i \leq n+1$):
\[
\begin{array}[t]{r@{\ }c@{\ }l}
\cn{1}
& ::= &
\mtcont
\Mid
\argcont{\cn{1}}{(\term,\env)}
\Mid
\succcont{\cn{1}}
\Mid
\appcont{\val}{\cn{1}}
\\
\cn{i}
& ::= &
\mtcont
\Mid
\cons{\cn{i}}{\cn{i-1}}
\end{array}
\]

\itemsp{ 
Environments:
$
\begin{array}[t]{r@{\ }c@{\ }l}
\env
& ::= &
\mtenv
\Mid
\extend{\env}{\var}{\val}
\end{array}
$}
\item Initial transition, transition rules, and final transition:
  see Figure~\ref{fig:abstract-machine-env-level-n-ctd}
\end{itemize}
\vspace{-0.2cm}
\caption{An environment-based abstract machine for the CPS hierarchy
  at level $n$}
\label{fig:abstract-machine-env-level-n}
\vspace{2mm}
\hrule
\end{figure}

\begin{sidewaysfigure}
\hrule
\vspace{2mm}
\begin{itemize}
\item
Initial transition, transition rules, and final transition ($1\leq
i \leq n,\;2 \leq j \leq n$):
\end{itemize}
\vspace{0.2cm}
\[
\hspace{-0.5cm}
\begin{machinetable}
\spacearound
\vspace{2mm}
\transition{\term}
           {\namedsixtuple{\evalconfname}
                          {\term}
                          {\mtenv}
                          {\mtcont}
                          {\mtcont}
                          {...}%
                          {\mtcont}
                         }
\spacearound
\transition{\namedsixtuple{\evalconfname}
                          {\synnum{m}}
                          {\env}
                          {\cn{1}}
                          {\cn{2}}
                          {...}%
                          {\cn{n+1}}
                         }
           {\namedquintuple{\contoneconfname}
                           {\cn{1}}
                           {\num{m}}
                           {\cn{2}}
                           {...}%
                           {\cn{n+1}}
                         }
\spacearound
\transition{\namedsixtuple{\evalconfname}
                           {\var}
                           {\env}
                           {\cn{1}}
                           {\cn{2}}
                           {...}%
                           {\cn{n+1}}
                         }
           {\namedquintuple{\contoneconfname}
                           {\cn{1}}
                           {\env\,(\var)}
                           {\cn{2}}
                           {...}%
                           {\cn{n+1}}
                         }
\spacearound
\transition{\namedsixtuple{\evalconfname}
                          {\lam{\var}
                               {\term}
                             }
                          {\env}
                          {\cn{1}}
                          {\cn{2}}
                          {...}%
                          {\cn{n+1}}
                        }
           {\namedquintuple{\contoneconfname}
                           {\cn{1}}
                           {\closure{\var}
                                    {\term}
                                    {\env}
                                  }
                           {\cn{2}}
                           {...}%
                           {\cn{n+1}}
                         }
\spacearound
\transition{\namedsixtuple{\evalconfname}
                          {\termzero\,\termone}
                          {\env}
                          {\cn{1}}
                          {\cn{2}}
                          {...}%
                          {\cn{n+1}}
                        }
           {\namedsixtuple{\evalconfname}
                          {\termzero}
                          {\env}
                          {\argcont{\cn{1}}
                                   {(\termone,\env)}
                          }
                          {\cn{2}}
                          {...}%
                          {\cn{n+1}}
                        } 
\spacearound
\transition{\namedsixtuple{\evalconfname}
                          {\synsucc{\term}}
                          {\env}
                          {\cn{1}}
                          {\cn{2}}
                          {...}%
                          {\cn{n+1}}
                        }  
           {\namedsixtuple{\evalconfname}
                          {\term}
                          {\env}
                          {\succcont{\cont}}
                          {\cn{2}}
                          {...}%
                          {\cn{n+1}}
                        }
\spacearound
\transition{\namedsixtuple{\evalconfname}
                          {\synresetn{i}{\term}}
                          {\env}
                          {\cn{1}}
                          {\cn{2}}
                          {...}%
                          {\cn{n+1}}
                        }
           {\namedeighttuple{\evalconfname}
                            {\term}
                            {\env}
                            {\mtcont}
                            {...}%
                            {\mtcont}
                            {\cons{\cn{i+1}}{(...(\cons{\cn{2}}{\cn{1}})...)}}
                            {\cn{i+2},\,...}
                            {\cn{n+1}}
                         }
\vspace{2mm}
\spacearound
\transition{\namedsixtuple{\evalconfname}
                          {\synshiftn{i}{\vark}{\term}}
                          {\env}
                          {\cn{1}}
                          {\cn{2}}
                          {...}%
                          {\cn{n+1}}
                        }
           {\namedeighttuple{\evalconfname}
                            {\term}
                            {\extend{\env}
                                    {\vark}
                                    {\cons{\cn{i}}{(...(\cons{\cn{2}}{\cn{1}})...)}}}
                            {\mtcont}
                            {...}%
                            {\mtcont}
                            {\cn{i+1}}
                            {...}%
                            {\cn{n+1}}
                          }
\spacearound
\transition{\namedquintuple{\contoneconfname}
                           {\mtcont}
                           {\val}
                           {\cn{2}}
                           {...}%
                           {\cn{n+1}}
                         }
           {\namedquintuple{\conttwoconfname}
                           {\cn{2}}
                           {\val}
                           {\cn{3}}
                           {...}%
                           {\cn{n+1}}
                         }
\spacearound
\transition{\namedquintuple{\contoneconfname}
                           {\argcont{\cont}
                                    {(\term,\env)}
                                  }
                           {\val}
                           {\cn{2}}
                           {...}%
                           {\cn{n+1}}
                         }
           {\namedsixtuple{\evalconfname}
                          {\term}
                          {\env}
                          {\appcont{\val}
                                   {\cn{1}}
                                 }
                          {\cn{2}}
                          {...}%
                          {\cn{n+1}}
                        }
\spacearound
\transition{\namedquintuple{\contoneconfname}
                           {\succcont{\cn{1}}}
                           {\num{m}}
                           {\cn{2}}
                           {...}%
                           {\cn{n+1}}
                         }
           {\namedquintuple{\contoneconfname}
                           {\cn{1}}
                           {\add{\num{m}}{1}}
                           {\cn{2}}
                           {...}%
                           {\cn{n+1}}
                         }
\spacearound
\transition{\namedquintuple{\contoneconfname}
                           {\appcont{\closure{\var}
                                             {\term}
                                             {\env}
                                           }
                                    {\cont}
                                  }
                           {\val}
                           {\cn{2}}
                           {...}%
                           {\cn{n+1}}
                         }
           {\namedsixtuple{\evalconfname}
                          {\term}
                          {\extend{\env}
                                  {\var}
                                  {\val}
                                }
                          {\cn{1}}
                          {\cn{2}}
                          {...}%
                          {\cn{n+1}}
                        }
\spacearound
\vspace{2mm}
\transition{\namedquintuple{\contoneconfname}
                           {\appcont{\cons{\cn{i}'}{(...(\cons{\cn{2}'}{\cn{1}'})...)}}
                                    {\cn{1}}
                                 }
                           {\val}
                           {\cn{2}}
                           {...}%
                           {\cn{n+1}}
                         }
           {\namedeighttuple{\contoneconfname}
                            {\cn{1}'}
                            {\val}
                            {\cn{2}',\,...}
                            {\cn{i}'}
                            {\cons{\cn{i+1}}{(...(\cons{\cn{2}}{\cn{1}})...)}}
                            {\cn{i+2}}
                            {...}%
                            {\cn{n+1}}
                          }
\spacearound
\transition{\namedquintuple{\conticonfname{j}}
                           {\mtcont}
                           {\val}
                           {\cn{j+1}}
                           {...}%
                           {\cn{n+1}}
                         }
           {\namedquintuple{\conticonfname{j+1}}
                           {\cn{j+1}}
                           {\val}
                           {\cn{j+2}}
                           {...}%
                           {\cn{n+1}}
                         }
\vspace{2mm}
\spacearound
\transition{\namedquintuple{\conticonfname{j}}
                           {\cons{\cn{j}}{(...(\cons{\cn{2}}{\cn{1}})...)}}
                           {\val}
                           {\cn{j+1}}
                           {...}%
                           {\cn{n+1}}
                         }
           {\namedquintuple{\conticonfname{1}}
                           {\cn{1}}
                           {\val}
                           {\cn{2}}
                           {...}%
                           {\cn{n+1}}
                         }
\vspace{2mm}
\spacearound
\transition{\namedpair{\conticonfname{n+1}}
                      {\cons{\cn{n+1}}{(...(\cons{\cn{2}}{\cn{1}})...)}}
                      {\val}
                    }
           {\namedquintuple{\conticonfname{1}}
                           {\cn{1}}
                           {\val}
                           {\cn{2}}
                           {...}%
                           {\cn{n+1}}
                         }
\vspace{2mm}
\spacearound
\transition{\namedpair{\conticonfname{n+1}}
                      {\mtcont}
                      {\val}
                    }
           {\val}
\end{machinetable}
\]
\vspace{-0.5cm}
\caption{An environment-based abstract machine for
  the CPS hierarchy at level $n$, ctd.}
\label{fig:abstract-machine-env-level-n-ctd}
\vspace{2mm}
\hrule
\end{sidewaysfigure}

\begin{figure}[b!!]
\hrule
\vspace{2mm}
\begin{itemize}
\item
Terms and values ($1 \leq i \leq n$):
$
\begin{array}[t]{r@{\ }c@{\ }l}
\term 
& ::= & 
\val
\Mid
\var
\Mid 
\synapp{\termzero}{\termone}
\Mid
\synsucc{\term}
\Mid
\synresetn{i}{\term}
\Mid
\synshiftn{i}{\vark}{\term}
\\[1mm]
\val
& ::= &
\synnum{m}
\Mid
\synlam{\var}{\term}
\Mid
\cn{i}
\end{array}
$
\item
Evaluation contexts ($2 \leq i \leq n+1$):
\[
\begin{array}[t]{r@{\ }c@{\ }l}
\cn{1}
& ::= &
\mtcont
\Mid
\argcont{\cn{1}}{\term}
\Mid
\succcont{\cn{1}}
\Mid
\appcont{\val}{\cn{1}}
\\
\cn{i}
& ::= &
\mtcont
\Mid
\cons{\cn{i}}{\cn{i-1}}
\end{array}
\]

\item Initial transition, transition rules, and final transition:
  see Figure~\ref{fig:abstract-machine-subst-level-n-ctd}
\end{itemize}
\vspace{-0.2cm}
\caption{A substitution-based abstract machine for the 
  CPS hierarchy at level $n$}
\label{fig:abstract-machine-subst-level-n}
\vspace{2mm}
\hrule
\end{figure}

\begin{sidewaysfigure}
\hrule
\vspace{2mm}
\begin{itemize}
\item
Initial transition, transition rules, and final transition ($1\leq
i \leq n,\;2 \leq j \leq n$):
\end{itemize}
\vspace{0.4cm}
\[
\hspace{-0.5cm}
\begin{machinetable}

\vspace{3mm}
\spacearound
\transition{\term}
           {\namedquintuple{\evalconfname}
                           {\term}
                           {\mtcont}
                           {\mtcont}
                           {...}%
                           {\mtcont}
                         }
\spacearound
\transition{\namedquintuple{\evalconfname}
                           {\synnum{m}}
                           {\cn{1}}
                           {\cn{2}}
                           {...}%
                           {\cn{n+1}}
                         }
           {\namedquintuple{\contoneconfname}
                           {\cn{1}}
                           {\synnum{m}}
                           {\cn{2}}
                           {...}%
                           {\cn{n+1}}
                         }
\spacearound
\transition{\namedquintuple{\evalconfname}
                           {\lam{\var}
                                {\term}
                              }
                           {\cn{1}}
                           {\cn{2}}
                           {...}%
                           {\cn{n+1}}
                         }
           {\namedquintuple{\contoneconfname}
                           {\cn{1}}
                           {\lam{\var}
                                {\term}}
                           {\cn{2}}
                           {...}%
                           {\cn{n+1}}
                         }
\spacearound
\transition{\namedquintuple{\evalconfname}
                           {\cn{i}'}
                           {\cn{1}}
                           {\cn{2}}
                           {...}%
                           {\cn{n+1}}
                         }
           {\namedquintuple{\contoneconfname}
                           {\cn{1}}
                           {\cn{i}'}
                           {\cn{2}}
                           {...}%
                           {\cn{n+1}}
                         }
\spacearound
\transition{\namedquintuple{\evalconfname}
                           {\synapp{\termzero}{\termone}}
                           {\cn{1}}
                           {\cn{2}}
                           {...}%
                           {\cn{n+1}}
                         }
           {\namedquintuple{\evalconfname}
                           {\termzero}
                           {\argcont{\cn{1}}
                                    {\pr{\termone}{\env}}}
                           {\cn{2}}
                           {...}%
                           {\cn{n+1}}
                         } 
\spacearound
\transition{\namedquintuple{\evalconfname}
                           {\synsucc{\term}}
                           {\cn{1}}
                           {\cn{2}}
                           {...}%
                           {\cn{n+1}}
                         }  
           {\namedquintuple{\evalconfname}
                           {\term}
                           {\succcont{\cont}}
                           {\cn{2}}
                           {...}%
                           {\cn{n+1}}
                         }
\spacearound
\transition{\namedquintuple{\evalconfname}
                           {\synresetn{i}{\term}}
                           {\cn{1}}
                           {\cn{2}}
                           {...}%
                           {\cn{n+1}}
                         }
           {\namedseventuple{\evalconfname}
                            {\term}
                            {\mtcont}
                            {...}%
                            {\mtcont}
                            {\cons{\cn{i+1}}{(...(\cons{\cn{2}}{\cn{1}})...)}}
                            {\cn{i+2},\,...}
                            {\cn{n+1}}
                          }
\vspace{3mm}
\spacearound
\transition{\namedquintuple{\evalconfname}
                           {\synshiftn{i}{\vark}{\term}}
                           {\cn{1}}
                           {\cn{2}}
                           {...}%
                           {\cn{n+1}}
                         }
           {\namedseventuple{\evalconfname}
                            {\subst{\term}
                                   {\vark}
                                   {\cons{\cn{i}}{(...(\cons{\cn{2}}{\cn{1}})...)}}}
                            {\mtcont}
                            {...}%
                            {\mtcont}
                            {\cn{i+1}}
                            {...}%
                            {\cn{n+1}}
                          }
\spacearound
\transition{\namedquintuple{\contoneconfname}
                           {\mtcont}
                           {\val}
                           {\cn{2}}
                           {...}%
                           {\cn{n+1}}
                         }
           {\namedquintuple{\conttwoconfname}
                           {\cn{2}}
                           {\val}
                           {\cn{3}}
                           {...}%
                           {\cn{n+1}}
                         }
\spacearound
\transition{\namedquintuple{\contoneconfname}
                           {\argcont{\cont}
                                    {\term}
                                  }
                           {\val}
                           {\cn{2}}
                           {...}%
                           {\cn{n+1}}
                         }
           {\namedquintuple{\evalconfname}
                           {\term}
                           {\appcont{\val}
                                    {\cn{1}}
                                  }
                           {\cn{2}}
                           {...}%
                           {\cn{n+1}}
                         }
\spacearound
\transition{\namedquintuple{\contoneconfname}
                           {\succcont{\cn{1}}}
                           {\synnum{m}}
                           {\cn{2}}
                           {...}%
                           {\cn{n+1}}
                         }
           {\namedquintuple{\contoneconfname}
                           {\cn{1}}
                           {\synnum{\add{\num{m}}{1}}}
                           {\cn{2}}
                           {...}%
                           {\cn{n+1}}
                         }
\spacearound
\transition{\namedquintuple{\contoneconfname}
                           {\appcont{(\synlam{\var}{\term})}
                                    {\cont}}
                           {\val}
                           {\cn{2}}
                           {...}%
                           {\cn{n+1}}
                         }
           {\namedquintuple{\evalconfname}
                           {\subst{\term}
                                  {\var}
                                  {\val}}
                           {\cn{1}}
                           {\cn{2}}
                           {...}%
                           {\cn{n+1}}
                         }
\vspace{3mm}
\spacearound
\transition{\namedquintuple{\contoneconfname}
                           {\appcont{\cons{\cn{i}'}{(...(\cons{\cn{2}'}{\cn{1}'})...)}}
                                    {\cn{1}}
                                  }
                           {\val}
                           {\cn{2}}
                           {...}%
                           {\cn{n+1}}
                         }
           {\namedeighttuple{\contoneconfname}
                            {\cn{1}'}
                            {\val}
                            {\cn{2}',\,...}
                            {\cn{i}'}
                            {\cons{\cn{i+1}}{(...(\cons{\cn{2}}{\cn{1}})...)}}
                            {\cn{i+2}}
                            {...}%
                            {\cn{n+1}}
                          }
\spacearound
\transition{\namedquintuple{\conticonfname{j}}
                           {\mtcont}
                           {\val}
                           {\cn{j+1}}
                           {...}%
                           {\cn{n+1}}
                         }
           {\namedquintuple{\conticonfname{j+1}}
                           {\cn{j+1}}
                           {\val}
                           {\cn{j+2}}
                           {...}%
                           {\cn{n+1}}
                         }
\vspace{3mm}
\spacearound
\transition{\namedquintuple{\conticonfname{j}}
                           {\cons{\cn{j}}{(...(\cons{\cn{2}}{\cn{1}})...)}}
                           {\val}
                           {\cn{j+1}}
                           {...}%
                           {\cn{n+1}}
                         }
           {\namedquintuple{\conticonfname{1}}
                           {\cn{1}}
                           {\val}
                           {\cn{2}}
                           {...}%
                           {\cn{n+1}}
                         }
\vspace{3mm}
\spacearound
\transition{\namedpair{\conticonfname{n+1}}
                      {\cons{\cn{n+1}}{(...(\cons{\cn{2}}{\cn{1}})...)}}
                      {\val}
                    }
           {\namedquintuple{\conticonfname{1}}
                           {\cn{1}}
                           {\val}
                           {\cn{2}}
                           {...}%
                           {\cn{n+1}}
                         }
\spacearound
\transition{\namedpair{\conticonfname{n+1}}
                      {\mtcont}
                      {\val}
                    }
           {\val}
\end{machinetable}
\]
\caption{A substitution-based abstract machine for the 
  the CPS hierarchy at level $n$, ctd.}
\label{fig:abstract-machine-subst-level-n-ctd}
\vspace{2mm}
\hrule
\end{sidewaysfigure}

\subsection{An environment-based
evaluator}
\label{subsec:env-interpreter-hierarchy}

At the $n$th level of the hierarchy, 
the language is extended with operators shift$_i$ and reset$_i$ for all
$i$ such that $1\leq i\leq n$.  The
evaluator for this
language is shown in Figures~\ref{fig:def-interpreter-level-n} and
\ref{fig:def-interpreter-level-n-ctd}.  If $n=1$, it coincides with the
evaluator displayed in Figure~\ref{fig:def-interpreter-level-one}.

The
evaluator uses $n+1$ layers of continuations.  In the five first clauses
(literal, variable, $\lambda$-abstraction, function application, and
application of the successor function), the continuations $k_2, \ldots,
k_{n+1}$ are passive: if the
evaluator were curried, they could be eta-reduced.  In the clauses
defining shift$_i$ and reset$_i$, the continuations $k_{i+2}, \ldots,
k_{n+1}$ are also passive.  Each pair of control operators is indexed by
the corresponding level in the hierarchy: reset$_i$ is used to ``push''
each successive continuation up to level $i$ onto level $i+1$ and to
reinitialize them with $\theta_1, \ldots, \theta_i$, which are the
successive CPS counterparts of the identity function; shift$_i$ is used
to abstract control up to level $i$ into a delimited continuation and to
reinitialize the successive continuations up to level $i$ with
$\theta_1, \ldots, \theta_i$.

Applying a delimited continuation that was abstracted up to level $i$
``pushes'' each successive continuation up to level $i$ onto level $i+1$
and restores the successive continuations that were captured in a
delimited continuation.
When such a delimited continuation completes, and when an expression
delimited by reset$_i$ completes, the successive continuations that were
pushed onto level $i+1$ are ``popped'' back into place and the
computation proceeds.

\subsection{An environment-based abstract machine}
\label{subsec:env-abstract-machine-hierarchy}

Defunctionalizing the
evaluator of Figures~\ref{fig:def-interpreter-level-n} and
\ref{fig:def-interpreter-level-n-ctd} yields the environment-based
abstract machine displayed in
Figures~\ref{fig:abstract-machine-env-level-n} and
\ref{fig:abstract-machine-env-level-n-ctd}.  If $n=1$, it coincides with
the abstract machine displayed in Figure~\ref{fig:abstract-machine-env}.

The abstract machine consists of $n+2$ sets of transitions:
$\mathit{eval}$ for interpreting terms and $\mathit{cont}_1, \ldots,
\mathit{cont}_{n+1}$ for interpreting the successive defunctionalized
continuations.  The set of possible values includes integers, closures
and captured contexts.

This abstract machine is an extension of the abstract machine displayed
in Figure~\ref{fig:abstract-machine-env} with $n+1$ contexts instead of 2
and the corresponding
transitions for shift$_i$ and reset$_i$.  Each meta$_{i+1}$-context
intervenes to process reset$_i$ expressions and to apply captured
continuations.  Otherwise, the successive contexts are passively threaded
to process literals, variables, $\lambda$-abstractions, function
applications, and applications of the successor function.

Given an environment $e$ and a series of successive contexts, a reset$_i$
expression $\synresetn{i}{t}$ is processed by evaluating $t$ with the
same environment $e$, $i$ empty contexts, and a meta$_{i+1}$-context over
which all the intermediate contexts have been pushed on.

\ %

Given an environment $e$ and a series of successive contexts, a shift
expression $\synshiftn{i}{k}{t}$ is processed by evaluating $t$ with an
extension of $e$ where $k$ denotes a composition of the $i$ surrounding
contexts, $i$ empty contexts, and the remaining outer contexts.  Applying
a captured context is achieved by pushing all the current contexts on the
next outer context, restoring the composition of the captured contexts,
and continuing with them.  Resuming a composition of captured contexts is
achieved by popping them off the next outer context and continuing with
them.

In order to relate the resulting abstract machine to the
evaluator, we define a partial function $\rawevalenvn$ mapping a term
$\term$ to a value $\val$ whenever the machine for level $n$, started
with
$\term$, stops
with $\val$.  The correctness of the machine with respect to the
evaluator is ensured by the following theorem:

\begin{theorem}
  {\em For any program
  $\term$ and any integer value $\num{m}$, $\namedone{\mainn}{\term} =
  \num{m}$ if and only if $\evalenvn{\term} = \num{m}$.
  \qed}
\end{theorem}

\subsection{A substitution-based abstract machine}
\label{subsec:subst-abstract-machine-hierarchy}

In the same fashion as in Section~\ref{subsec:abstract-machine-subst}, we
construct the substitution-based abstract machine corresponding to the
environment-based abstract machine of
Section~\ref{subsec:env-abstract-machine-hierarchy}.  The result is
displayed in Figures~\ref{fig:abstract-machine-subst-level-n} and
\ref{fig:abstract-machine-subst-level-n-ctd}.  If $n=1$, it coincides
with the abstract machine displayed in
Figure~\ref{fig:abstract-machine-subst}.

The $n$th level contains $n+1$ evaluation contexts and each context
$\ncont{i}$ can be viewed as a stack of non-empty contexts
$\ncont{i-1}$. Terms are decomposed as
\[
  \ncont{n+1}\ndel{n}\ncont{n}\ndel{n-1}\ncont{n-1}\ndel{n-2}\cdots\ndel{2}\metacont\ndel{1}\cont[\term],
\]

\noindent
where each $\ndel{i}$ represents
a context delimiter of level $i$.  All the control operators that
occur at the $j$th level (with $j<n$) of the hierarchy do not use the
contexts $j+2, \ldots, n+1$. The functions $\rawdecompose$ and its
inverse $\rawplug$ can be read off the machine, as for level 1.

The transitions of the machine for level $j$ are ``embedded'' in
the machine for level $j+1$; the extra components are threaded but
not used.  

We define a partial function $\rawevalsubn$ capturing the evaluation
by the substitution-based abstract machine for an arbitrary level $n$,
analogously to the definition of $\rawevalenvn$. Now we can relate
evaluation with the environment-based and the substitution-based
abstract machines for level $n$.

\begin{theorem}
  {\em For any program $\term$, either both $\evalsubn{\term}$ and
  $\evalenvn{\term}$ are undefined, or there exist values $\val,\val'$
  such that $\evalsubn{\term}=\val$, $\evalenvn{\term}=\val'$ and
  $\realn{\val'}=\val$.
  
  The definition of $\rawrealn$ extends that of $\rawreal$ from
  Theorem~\ref{thm:simulation-level-1-env-subst-am} in such a way that it
  is homomorphic for all the contexts $\ncont{i}$, with $2\leq i\leq
  n$.
  \qed}
\end{theorem}

\subsection{A reduction semantics}
\label{subsec:reduction-semantics-hierarchy}

Along the same lines as in Section~\ref{subsec:a-reduction-semantics}, we
construct the reduction semantics for the CPS hierarchy based on the
abstract machine of Figures~\ref{fig:abstract-machine-subst-level-n}
and~\ref{fig:abstract-machine-subst-level-n-ctd}.  For an arbitrary level
$n$ we obtain the following set of reduction rules, for all $1\leq i\leq
n$; they define the actual redexes:
\[
\begin{array}{ll}
(\delta)\;&
\decompn{1}{[\synsucc{\synnum{m}}]}
\redn
\decompn{1}{[\synnum{m+1}]}\\[2mm]
(\beta_{\lambda})\; &
\decompn{1}{[\synapp{\lamp{\var}{\term}}{\val}]}
\redn
\decompn{1}{[\subst{\term}{\var}{\val}]}
\\[2mm]
(\rawshift^i_{\lambda})\; &
\decompn{1}{[\synshiftn{i}{\vark}{\term}]} 
\redn\\
&\decompn{i+1}{\ndel{i}\mtcont\ldots\ndel{1}[\subst{\term}{\vark}{\consn{i}}]}
\\[2mm]
(\beta^i_{\mathit{ctx}})\; &
\decompn{1}{[\synapp{\consntwo{i}}{\val}]}  
\redn\\
&\ncont{n+1}\ndel{n}\cdots\ndel{i+1}
\cons{\ncont{i+1}}{(\ldots(\cons{\ncont{2}}{\ncont{1}})\ldots)}\ndel{i}\nconttwo{i}\ndel{i-1}\cdots\ndel{1}\nconttwo{1}[\val]

\\[2mm]
(\text{Reset}^i)\; &
 \decompn{1}{[\synresetn{i}{\val}]}  
\redn
 \decompn{1}{[\val]}
\end{array}
\]

Each level contains all the reductions from lower levels, and these
reductions are compatible with additional layers of evaluation
contexts.  In particular, at level $0$ there are only $\delta$- and
$\beta_\lambda$-reductions.

The values and evaluation contexts are already specified in the
abstract machine. Moreover, the 
potential
redexes are defined according to the
following grammar:
\begin{align*}
  \potredexn & ::= \synsucc{\val} 
  \Mid
  \synapp{\val_0}{\val_1} 
  \Mid 
  \synshiftn{i}{\vark}{\term}
  \Mid 
  \synresetn{i}{\val}\quad
  (1\leq i\leq n)
\end{align*}

\begin{lemma}[Unique decomposition for level $n$]
  A program \term\ is either a value or there exists a unique sequence
  of contexts $\ncont{1},\ldots,\ncont{n+1}$ and a potential redex
  $\potredexn$ such that\linebreak
  $\term=\plug{\decompn{1}{[\potredexn]}}$.
  \qed
\end{lemma}

Evaluating a term using either the derived reduction rules or the
substitution-based abstract machine from
Section~\ref{subsec:subst-abstract-machine-hierarchy} yields the same
result:

\begin{theorem}
  {\em For any 
  program \term\ and any value $\val$, $\evalsubn{\term} = \val$ if
  and only if $\term\redntrans\val$, where $\redntrans$ is the
  reflexive, transitive closure of $\redn$.
  \qed}
\end{theorem}

As in
Section~\ref{subsec:a-reduction-semantics}, using refocusing, one can go
from a given reduction semantics of
Section~\ref{subsec:reduction-semantics-hierarchy} into a pre-abstract
machine
and the corresponding eval/apply abstract machine of
Figures~\ref{fig:abstract-machine-subst-level-n} and
\ref{fig:abstract-machine-subst-level-n-ctd}.

\subsection{Beyond CPS}

As in Section~\ref{subsec:beyond-CPS}, one can define a family of concatenation
functions over contexts and use it to implement composable continuations
in the CPS hierarchy, giving rise to a family of control operators
$\mathcal{F}_n$ and $\#_n$.  Again the modified environment-based abstract
machine does not immediately correspond to a defunctionalized
continuation-passing evaluator.  Such control operators go beyond
traditional CPS.

\subsection{Static vs.~dynamic delimited continuations}

As in Section~\ref{subsec:static-vs-dynamic-del-conts}, one can
illustrate the difference between static and dynamic delimited
continuations in the CPS hierarchy.  For example, replacing
shift$_2$ and reset$_2$ respectively by
$\mathcal{F}_2$ and $\#_2$ in Danvy and Filinski's version of Abelson
and Sussman's generator of
triples~\cite[Section~3]{Danvy-Filinski:LFP90} yields a list in
reverse order.\footnote{Thanks are due to an anonymous reviewer for
  pointing this out.}

\subsection{Summary and conclusion}

We have generalized the results presented in
Section~\ref{sec:derivation} from level 1 to the whole CPS hierarchy
of control operators shift$_n$ and reset$_n$.  Starting from the
original
evaluator for the $\lambda$-calculus with shift$_n$ and reset$_n$ that
uses $n+1$ layers of continuations, we have derived two abstract
machines, an environment-based one and a substitution-based one; each of
these machines use $n+1$ layers of evaluation contexts.  Based on the
substitution-based machine we have obtained a reduction semantics for the
$\lambda$-calculus extended with shift$_n$ and reset$_n$ which, by
construction, is
sound with respect to CPS.

\section{Programming in the CPS hierarchy}
\label{sec:programming-in-hierarchy}

To finish, we present new examples of programming in the CPS hierarchy.
The examples are normalization functions.  In
Sections~\ref{subsec:normalization-by-evaluation} and
\ref{subsec:the-free-monoid}, we first describe normalization by
evaluation and we present the simple example of the free monoid.  In
Section~\ref{subsec:propositional-logic}, we present a function mapping a
proposition into its disjunctive normal form; this normalization function
uses delimited continuations.  In
Section~\ref{subsec:a-generalization-of-propositional-logic}, we
generalize the normalization functions of
Sections~\ref{subsec:the-free-monoid} and
\ref{subsec:propositional-logic} to a hierarchical language of
units and products, and we express the corresponding normalization
function in the CPS hierarchy.

\subsection{Normalization by evaluation}
\label{subsec:normalization-by-evaluation}

Normalization by evaluation is a `reduction-free' approach to
normalizing terms.  Instead of reducing a term to its normal form, one
evaluates this term into a non-standard model and reifies its
denotation into its normal form~\cite{Dybjer-Filinski:APPSEM00}:
\[
  \begin{array}{r@{\ }c@{\ }l}
  \mathit{eval} & : & \mathit{term} \rightarrowp \mathit{value}
  \\
  \mathit{reify} & : & \mathit{value} \rightarrowp \mathit{term}^{\mathrm{nf}}
  \\
  \mathit{normalize} & : & \mathit{term} \rightarrowp \mathit{term}^{\mathrm{nf}}
  \\
  \mathit{normalize} & = & \syncompose{\mathit{reify}}{\mathit{eval}}
  \end{array}
\]

Normalization by evaluation has been developed in
intuitionistic type
theory~\cite{Coquand-Dybjer:MSCS97,Martin-Loef:SLS73}, proof
theory~\cite{Berger-al:NADA98,Berger-Schwichtenberg:LICS91}, category
theory~\cite{Altenkirch-Hofmann-Streicher:CTCS95}, and partial
evaluation~\cite{Danvy:POPL96,Danvy:PEPT98}, where it has emerged as a
new field of application for delimited
continuations~\cite{%
Balat-Danvy:GPCE02,%
Danvy:PEPT98,%
Dybjer-Filinski:APPSEM00,%
Filinski:TLCA01,%
Grobauer-Yang:HOSC01,%
Helsen-Thiemann:ASIAN98,%
Sumii-Kobayashi:HOSC01%
}.

\subsection{The free monoid}
\label{subsec:the-free-monoid}

A source term in the free monoid is either a variable, the unit element,
or the product of two terms:
\[
\begin{array}[t]{r@{\hspace{2mm}}c@{\hspace{2mm}}l}
\synprop \; \ni \; \synvarterm
& ::= &
\synvar{x} \Mid
\varepsilon \Mid
\synvarterm \star \synvarterm'
\end{array}
\]

\noindent
The product is associative and the unit element is neutral.  These
properties justify the following conversion rules:
\begin{eqnarray*}
\synvarterm \star (\synvarterm' \star \synvarterm'')
& \leftrightarrow &
(\synvarterm \star \synvarterm') \star \synvarterm''
\\
\synvarterm \star \varepsilon
& \leftrightarrow &
\synvarterm
\\
\varepsilon \star \synvarterm
& \leftrightarrow &
\synvarterm
\end{eqnarray*}

We aim (for example) for list-like flat normal forms:
\[
\begin{array}[t]{r@{\hspace{2mm}}c@{\hspace{2mm}}l}
\dnfprop \; \ni \; \widehat{\synvarterm}
& ::= &
\varepsilon^{\mathrm{nf}} \Mid x \star^{\mathrm{nf}} \widehat{\synvarterm}
\end{array}
\]

In a reduction-based approach to normalization, one would orient the
conversion rules into reduction rules and one would apply these reduction
rules until a normal form is obtained:
\begin{eqnarray*}
\synvarterm \star (\synvarterm' \star \synvarterm'')
& \leftarrow &
(\synvarterm \star \synvarterm') \star \synvarterm''
\\
\varepsilon \star \synvarterm
& \rightarrow &
\synvarterm
\end{eqnarray*}

In a reduction-free approach to normalization, one defines a
normalization function as the composition of a non-standard evaluation
function and a reification function.  Let us state such a normalization
function.

The non-standard domain of values is the transformer
\[
  \begin{array}{r@{\ }c@{\ }l}
  \mathit{value}
  & = &
  \dnfprop \rightarrowp \dnfprop.
  \end{array}
\]

The evaluation function is defined by induction over the syntax of source
terms, and the reification function inverts it:
\[
  \begin{array}[t]{r@{\ }c@{\ }l}
  \mathit{eval}\:x
  & = &
  \synlam{\synvarterm}{x \star^{\mathrm{nf}}\mbox{\hspace{-0.2mm}}\synvarterm}
  \\
  \mathit{eval}\:\varepsilon
  & = &
  \synlam{\synvarterm}{\synvarterm}
  \\
  \mathit{eval}\:(\synvarterm \star \synvarterm')
  & = &
  (\mathit{eval}\:\synvarterm) \circ (\mathit{eval}\:\synvarterm')
  \\[1ex]
  \mathit{reify}\:v
  & = &
  \synapp{v}{\varepsilon^{\mathrm{nf}}}
  \\[1ex]
  \mathit{normalize}\:\synvarterm
  & = &
  \mathit{reify}\:(\mathit{eval}\:\synvarterm)
  \end{array}
\]

In effect, $\mathit{eval}$ is a 
homomorphism from the source monoid to the
monoid of transformers (unit is mapped to unit and products are mapped to
products) and the normalization function hinges on the built-in
associativity of function composition.  Beylin, Dybjer, Coquand, and Kinoshita have studied its
theoretical
content~\cite{Beylin-Dybjer:TYPES95,Coquand-Dybjer:MSCS97,Kinoshita:MJ98}.
From a (functional) programming standpoint, the reduction-based approach
amounts to flattening a tree iteratively by reordering it, and the
reduction-free approach amounts to flattening a tree with an accumulator.

\subsection{A language of propositions}
\label{subsec:propositional-logic}

A source term, \ie, a proposition, is either a variable, a literal (true
or false), a conjunction, or a disjunction:
\[
\begin{array}[t]{r@{\hspace{2mm}}c@{\hspace{2mm}}l}
\synprop \; \ni \; \synvarterm
& ::= &
\synvar{x} \Mid
\syntrue \Mid
\synconj{\synvarterm}{\synvarterm'} \Mid
\synfalse \Mid
\syndisj{\synvarterm}{\synvarterm'}
\end{array}
\]

\noindent
Conjunction and disjunction are associative and distribute over each
other; $\syntrue$ is neutral for conjunction and absorbant for
disjunction; and $\synfalse$ is neutral for disjunction and absorbant for
conjunction.

We aim (for example) for list-like disjunctive normal forms:
\[
\begin{array}[t]{r@{\hspace{2mm}}c@{\hspace{2mm}}l}
\dnfprop \; \ni \; \widehat{\synvarterm} & ::= & d
\\
\dnfpropd \; \ni \; d & ::= & \dnffalse \Mid \dnfdisj{c}{d}
\\
\dnfpropc \; \ni \; c & ::= & \dnftrue \Mid \dnfconj{\synvar{x}}{c}
\end{array}
\]

\noindent
Our
normalization function is the result of composing a non-standard
evaluation function and a reification function.  We state them below
without proof.

Given the domains of transformers
\[
  \begin{array}{rcl}
  \Transformer{1}
  & = &
  \dnfpropc \rightarrowp \dnfpropc
  \\
  \Transformer{2}
  & = &
  \dnfpropd \rightarrowp \dnfpropd
  \end{array}
\]

\noindent
the non-standard domain of values is $\ansn{1}$, where
\[
  \begin{array}{rcl}
  \ansn{2}
  & = &
  \Transformer{2}
  \\
  \ansn{1}
  & = &
  (\Transformer{1} \rightarrowp \ansn{2}) \rightarrowp \ansn{2}.
  \end{array}
\]

\noindent
The evaluation function is defined by induction over the syntax of
source terms, and the reification function inverts it:
\[
\begin{array}{r@{\hspace{2mm}}c@{\hspace{2mm}}l}
&&
\\%
\synapptwo{\evalpc{\synvar{x}}}{k}{d}
& = &
               {\synapptwo{k}
                          {\synlamp{c}
                                   {\dnfconj{\synvar{x}}
                                            {c}}}
                          {d}}
\\
\synapptwo{\evalpc{\syntrue}}{k}{d}
& = &
               {\synapptwo{k}
                          {\synlamp{c}
                                   {c}}
                          {d}}
\\
\synapptwo{\evalpc{\synconjp{\synvarterm}{\synvarterm'}}}{k}{d}
& = &
               {\synapptwo{\evalpc{\synvarterm}}
                          {\synlamp{\transformer{1}}
                                   {\synapp{\evalpc{\synvarterm'}}
                                           {\synlamp{\transformer{1}'}
                                                   {\synapp{k}
                                                           {\syncomposep{\transformer{1}}
                                                                        {\transformer{1}'}}}}}}
                          {d}}
\\
\synapptwo{\evalpc{\synfalse}}{k}{d}
& = &
               {d}
\\
\synapptwo{\evalpc{\syndisjp{\synvarterm}{\synvarterm'}}}{k}{d}
& = &
               {\synapp{\synapp{\evalpc{\synvarterm}}
                               {k}}
                       {\synapptwop{\evalpc{\synvarterm'}}
                                           {k}
                                           {d}}}
\\[1.5ex]
\reifyn{0}{v}
& = &
\synapptwo{v}
          {\synlamp{\transformer{1}}
                   {\synlam{d}
                           {\dnfdisj{\synappp{\transformer{1}}
                                             {\dnftrue}}
                                    {d}}}}
          {\dnffalse}
\\[1.5ex]
\norm{\synvarterm}
& = &
\reifyn{0}{\evalpcp{\synvarterm}}
\end{array}
\]

\noindent
This normalization function uses a continuation $k$, an accumulator $d$
to flatten disjunctions, and another one $c$ to flatten conjunctions.
The continuation is delimited: the three first clauses of $\rawevalpc$
are in CPS; in the fourth, $k$ is discarded (accounting for the fact that
$\synfalse$ is absorbant for conjunction); and in the last, $k$ is
duplicated and used in non-tail position (achieving the distribution of
conjunctions over disjunctions).  The continuation and the accumulators
are initialized in the definition of $\rawreify_0$.

Uncurrying the continuation and mapping $\rawevalpc$ and $\rawreify_0$ back
to direct style yield the following definition, which lives at level 1 of
the CPS hierarchy:
\[
\begin{array}[t]{r@{\hspace{2mm}}c@{\hspace{2mm}}l}
\evalpd{\synvar{x}}{d}
& = &
\synpair{\synlam{c}
                {\dnfconj{x}
                         {c}}}
        {d}
\\
\evalpd{\syntrue}{d}
& = &
\synpair{\synlam{c}
                {c}}
        {d}
\\
\evalpd{\synconjp{\synvarterm}{\synvarterm'}}{d}
& = &
\vsynlet{\synpair{\transformer{1}}{d}}
        {\evalpd{\synvarterm}{d}}
        {\vsynlet{\synpair{\transformer{1}'}{d}}
                 {\evalpd{\synvarterm'}{d}}
                 {\synpair{\syncompose{\transformer{1}}{\transformer{1}'}}{d}}}
\\
\evalpd{\synfalse}{d}
& = &
\synshift{k}
         {d}
\\
\evalpd{\syndisjp{\synvarterm}{\synvarterm'}}{d}
& = &
\synshift{k}
         {\synapp{k}
           {\evalpdp{\synvarterm}
             {\synreset{\synapp{k}
                 {\evalpdp{\synvarterm'}
                   {d}}}}}}
\\[1.5ex]
\reifyn{1}{v}
& = &
\synresetvlet{\synpair{\transformer{1}}
                      {d}}
             {\synapp{v}
                     {\dnffalse}}
             {\dnfdisj{\synappp{\transformer{1}}
                               {\dnftrue}}
                      {d}}
\\[3.5ex]
\norm{\synvarterm}
& = &
\reifyn{1}{\evalpdonep{\synvarterm}}
\end{array}
\]

The three first clauses of $\rawevalpd$ are in direct style; the two
others abstract control with shift.  In the fourth clause, the context is
discarded; and in the last clause, the context is duplicated and
composed.  The context and the accumulators are initialized in the
definition of $\rawreify_1$.

This direct-style version makes it even more clear than the CPS
version that the accumulator for the disjunctions in normal form is a
threaded state.  A continuation-based, state-based version (or better,
a monad-based one) can therefore be written---but it is out of scope
here.

\subsection{A hierarchical language of units and products}
\label{subsec:a-generalization-of-propositional-logic}

We consider a generalization of propositional logic where a source
term is either a variable, a unit in a hierarchy of units, or a
product in a hierarchy of products:
\[
\begin{array}[t]{r@{\ }c@{\ }l}
\rawsynmon \; \ni \; \synvarterm
& ::= &
\synvar{x} \Mid
\synunit{i} \Mid
\synprod{i}{\synvarterm}{\synvarterm'}
\\
&&
\mathit{where}\; 1 \leq i \leq n.
\end{array}
\]

\noindent
All the products are associative.
All units are
neutral for products with the same index.

\begin{description}
  
\item[The free monoid]
  The language corresponds to that of the free monoid if $n=1$, as in
  Section~\ref{subsec:the-free-monoid}.
\vspace{1mm}  
\item[Boolean logic]
  The language corresponds to that of propositions if $n=2$, as in
  Section~\ref{subsec:propositional-logic}: $\synunit{1}$ is $\syntrue$,
  $\rawsynprod{1}$ is $\rawsynconj$, $\synunit{2}$ is $\synfalse$, and
  $\rawsynprod{2}$ is $\rawsyndisj$.
\vspace{1mm}  

\item[Multi-valued logic]
  In general, for each $n>2$ we can consider a suitable $n$-valued
  logic~\cite{Ginsberg:CI88}; for example, in case $n=4$, the language
  corresponds to that
  of Belnap's bilattice $\mathcal{FOUR}$~\cite{Belnap:CAP76}. It is also possible
  to modify the normalization function to work for less regular
  logical structures (\eg, other bilattices).
\vspace{1mm}  

\item[Monads]
  In general, the language corresponds to that of layered
  monads~\cite{Moggi:IaC91}: each unit element is the unit of the
  corresponding monad, and each product is the `bind' of the
  corresponding monad.  In practice, layered monads are collapsed into
  one for programming consumption~\cite{Filinski:POPL99}, but prior to
  this collapse, all the individual monad operations coexist in the
  computational soup.

\end{description}

In the remainder of this section, we assume that all the products,
besides being associative, distribute over each other, and that all
units, besides being neutral for products with the same index, are
absorbant for products with other indices.
We aim (for example) for a generalization of disjunctive normal forms:
\[
\begin{array}[t]{r@{\ }c@{\ }l}
\dnfmontop \; \ni \; \widehat{\synvarterm}{\phantom{_n}}
& ::= &
\synvartermi{n}
\\
\dnfmon{n} \; \ni \; \synvartermi{n}
& ::= &
\dnfunit{n} \Mid \dnfprod{n}{\synvartermi{n-1}}{\synvartermi{n}}
\\
& \vdots &                 
\\
\dnfmon{1} \; \ni \; \synvartermi{1}
& ::= &
\dnfunit{1} \Mid \dnfprod{1}{\synvartermi{0}}{\synvartermi{1}}
\\
\dnfmon{0} \; \ni \; \synvartermi{0}
& ::= &
\synvar{x}
\end{array}
\]

\noindent
For presentational reasons, in the remainder of this section we
arbitrarily fix $n$ to be 5.

Our normalization function is the result of composing a non-standard
evaluation function and a reification function.  We state them below
without proof.  Given the domains of transformers
\[
  \begin{array}{rcl}
  \Transformer{1}
  & = &
  \dnfmon{1} \rightarrowp \dnfmon{1}
  \\
  \Transformer{2}
  & = &
  \dnfmon{2} \rightarrowp \dnfmon{2}
  \\
  \Transformer{3}
  & = &
  \dnfmon{3} \rightarrowp \dnfmon{3}
  \\
  \Transformer{4}
  & = &
  \dnfmon{4} \rightarrowp \dnfmon{4}
  \\
  \Transformer{5}
  & = &
  \dnfmon{5} \rightarrowp \dnfmon{5}
  \end{array}
\]

\noindent
the non-standard domain of values is $\ansn{1}$, where

\[
  \begin{array}{rcl}
  \ansn{5}
  & = &
  \Transformer{5}
  \\
  \ansn{4}
  & = &
  (\Transformer{4} \rightarrowp \ansn{5}) \rightarrowp \ansn{5}
  \\
  \ansn{3}
  & = &
  (\Transformer{3} \rightarrowp \ansn{4}) \rightarrowp \ansn{4}
  \\
  \ansn{2}
  & = &
  (\Transformer{2} \rightarrowp \ansn{3}) \rightarrowp \ansn{3}
  \\
  \ansn{1}
  & = &
  (\Transformer{1} \rightarrowp \ansn{2}) \rightarrowp \ansn{2}.
  \end{array}
\]

\noindent
The evaluation function is defined by induction over the syntax of
source terms, and the reification function inverts it:
\[
\begin{array}[t]{r@{\hspace{2mm}}c@{\hspace{2mm}}l}
\synappsix{\evalfour{\synvar{x}}}{k_1}{k_2}{k_3}{k_4}{\synvartermi{5}}
& = &
\synappsix{k_1}{\synlamp{\synvartermi{1}}{\dnfprod{1}{\synvar{x}}{\synvartermi{1}}}}{k_2}{k_3}{k_4}{\synvartermi{5}}
\\
\synappsix{\evalfour{\synunit{1}}}{k_1}{k_2}{k_3}{k_4}{\synvartermi{5}}
& = &
\synappsix{k_1}{\synlamp{\synvartermi{1}}{\synvartermi{1}}}{k_2}{k_3}{k_4}{\synvartermi{5}}
\\
\synappsix{\evalfour{\synprodp{1}{\synvarterm}{\synvarterm'}}}{k_1}{k_2}{k_3}{k_4}{\synvartermi{5}}
& = &
\synappsix{\evalfour{\synvarterm}}
          {\synlamp{\transformer{1}}
                   {\synapp{\evalfour{\synvarterm'}}
                           {\synlamp{\transformer{1}'}
                                    {\synapp{k_1}
                                            {\syncomposep{\transformer{1}}
                                                         {\transformer{1}'}}}}}}
          {k_2}{k_3}{k_4}{\synvartermi{5}}
\\
\synappsix{\evalfour{\synunit{2}}}{k_1}{k_2}{k_3}{k_4}{\synvartermi{5}}
& = &
\synappfive{k_2}{\synlamp{\synvartermi{2}}{\synvartermi{2}}}{k_3}{k_4}{\synvartermi{5}}
\\
\synappsix{\evalfour{\synprodp{2}{\synvarterm}{\synvarterm'}}}{k_1}{k_2}{k_3}{k_4}{\synvartermi{5}}
& = &
\synappsix{\evalfour{\synvarterm}}
          {k_1}
          {\synlamp{\transformer{2}}
                   {\synappthree{\evalfour{\synvarterm'}}
                                {k_1}
                                {\synlamp{\transformer{2}'}
                                         {\synapp{k_2}
                                                 {\syncomposep{\transformer{2}}
                                                              {\transformer{2}'}}}}}}
          {k_3}{k_4}{\synvartermi{5}}
\\
\synappsix{\evalfour{\synunit{3}}}{k_1}{k_2}{k_3}{k_4}{\synvartermi{5}}
& = &
\synappfour{k_3}{\synlamp{\synvartermi{3}}{\synvartermi{3}}}{k_4}{\synvartermi{5}}
\\
\synappsix{\evalfour{\synprodp{3}{\synvarterm}{\synvarterm'}}}{k_1}{k_2}{k_3}{k_4}{\synvartermi{5}}
& = &
\synappsix{\evalfour{\synvarterm}}
          {k_1}
          {k_2}
          {\synlamp{\transformer{3}}
                   {\synappfour{\evalfour{\synvarterm'}}
                               {k_1}
                               {k_2}
                               {\synlamp{\transformer{3}'}
                                        {\synapp{k_3}
                                                {\syncomposep{\transformer{3}}
                                                             {\transformer{3}'}}}}}}
          {k_4}{\synvartermi{5}}
\\
\synappsix{\evalfour{\synunit{4}}}{k_1}{k_2}{k_3}{k_4}{\synvartermi{5}}
& = &
\synappthree{k_4}{\synlamp{\synvartermi{4}}{\synvartermi{4}}}{\synvartermi{5}}
\\
\synappsix{\evalfour{\synprodp{4}{\synvarterm}{\synvarterm'}}}{k_1}{k_2}{k_3}{k_4}{\synvartermi{5}}
& = &
\synappsix{\evalfour{\synvarterm}}
          {k_1}
          {k_2}
          {k_3}
          {\synlamp{\transformer{4}}
                   {\synappfive{\evalfour{\synvarterm'}}
                               {k_1}
                               {k_2}
                               {k_3}
                               {\synlamp{\transformer{4}'}
                                        {\synapp{k_4}
                                                {\syncomposep{\transformer{4}}
                                                             {\transformer{4}'}}}}}}
          {\synvartermi{5}}
\\
\synappsix{\evalfour{\synunit{5}}}{k_1}{k_2}{k_3}{k_4}{\synvartermi{5}}
& = &
\synvartermi{5}
\\
\synappsix{\evalfour{\synprodp{5}{\synvarterm}{\synvarterm'}}}{k_1}{k_2}{k_3}{k_4}{\synvartermi{5}}
& = &
\synappsix{\evalfour{\synvarterm}}
          {k_1}
          {k_2}
          {k_3}
          {k_4}
          {\synappsixp{\evalfour{\synvarterm'}}
                      {k_1}
                      {k_2}
                      {k_3}
                      {k_4}
                      {\synvartermi{5}}}
\\[1ex]
\reifyn{0}{v}
& = &
\synappsixv{v}
           {\synlamp{\transformer{1}}
                    {\synlam{k_2}
                            {\synapp{k_2}
                                    {\synlamp{\synvartermi{2}}
                                             {\dnfprod{2}
                                                      {\synappp{\transformer{1}}
                                                               {\dnfunit{1}}}
                                                      {\synvartermi{2}}}}}}}
           {\synlamp{\transformer{2}}
                    {\synlam{k_3}
                            {\synapp{k_3}
                                    {\synlamp{\synvartermi{3}}
                                             {\dnfprod{3}
                                                      {\synappp{\transformer{2}}
                                                               {\dnfunit{2}}}
                                                      {\synvartermi{3}}}}}}}
           {\synlamp{\transformer{3}}
                    {\synlam{k_4}
                            {\synapp{k_4}
                                    {\synlamp{\synvartermi{4}}
                                             {\dnfprod{4}
                                                      {\synappp{\transformer{3}}
                                                               {\dnfunit{3}}}
                                                      {\synvartermi{4}}}}}}}
           {\synlamp{\transformer{4}}
                    {\synlam{\synvartermi{5}}
                            {\dnfprod{5}
                                     {\synappp{\transformer{4}}
                                              {\dnfunit{4}}}
                                     {\synvartermi{5}}}}}
           {\synunit{5}}
 \\[6ex]
 \norm{\synvarterm}
 & = &
 \reifyn{0}{\evalfourp{\synvarterm}}
\end{array}
\]

\noindent
This normalization function uses four delimited continuations $k_1$,
$k_2$, $k_3$, $k_4$ and five accumulators $t_1$, $t_2$, $t_3$, $t_4$,
$t_5$ to flatten each of the successive products.  In the clause of each
$\synunit{i}$, the continuations $k_1, \ldots, k_{i-1}$ are discarded,
accounting for the fact that $\synunit{i}$ is absorbant for
$\rawsynprod{1}, \ldots, \rawsynprod{i-1}$, and the identity function is
passed to $k_i$, accounting for the fact that $\synunit{i}$ is neutral
for $\rawsynprod{i}$.  In the clause of each $\rawsynprod{i+1}$, the
continuations $k_1, \ldots, k_i$ are duplicated and used in non-tail
position, achieving the distribution of $\rawsynprod{i+1}$ over
$\rawsynprod{1}, \ldots, \rawsynprod{i}$.  The continuations and the
accumulators are initialized in the definition of $\rawreify_0$.

This normalization function lives at level 0 of the CPS hierarchy, but we
can express it at a higher level using shift and reset.  For example,
uncurrying $k_3$ and $k_4$ and mapping $\rawevalfour$ and $\rawreify_0$
back to direct style twice yield the following intermediate definition,
which lives at level 2:
\vspace{-0.5mm}
\[
\begin{array}{@{}r@{\hspace{2mm}}c@{\hspace{2mm}}l@{}}
\synappfour{\evaltwo{\synvar{x}}}{k_1}{k_2}{\synvartermi{5}}
& = &
\synappfour{k_1}{\synlamp{\synvartermi{1}}{\dnfprod{1}{\synvar{x}}{\synvartermi{1}}}}{k_2}{\synvartermi{5}}
\\
\synappfour{\evaltwo{\synunit{1}}}{k_1}{k_2}{\synvartermi{5}}
& = &
\synappfour{k_1}{\synlamp{\synvartermi{1}}{\synvartermi{1}}}{k_2}{\synvartermi{5}}
\\
\synappfour{\evaltwo{\synprodp{1}{\synvarterm}{\synvarterm'}}}{k_1}{k_2}{\synvartermi{5}}
& = &
\synappfour{\evaltwo{\synvarterm}}
          {\synlamp{\transformer{1}}
                   {\synapp{\evaltwo{\synvarterm'}}
                           {\synlamp{\transformer{1}'}
                                    {\synapp{k_1}
                                            {\syncomposep{\transformer{1}}
                                                         {\transformer{1}'}}}}}}
          {k_2}{\synvartermi{5}}
\\
\synappfour{\evaltwo{\synunit{2}}}{k_1}{k_2}{\synvartermi{5}}
& = &
\synappthree{k_2}{\synlamp{\synvartermi{2}}{\synvartermi{2}}}{\synvartermi{5}}
\\
\synappfour{\evaltwo{\synprodp{2}{\synvarterm}{\synvarterm'}}}{k_1}{k_2}{\synvartermi{5}}
& = &
\synappfour{\evaltwo{\synvarterm}}
          {k_1}
          {\synlamp{\transformer{2}}
                   {\synappthree{\evaltwo{\synvarterm'}}
                                {k_1}
                                {\synlamp{\transformer{2}'}
                                         {\synapp{k_2}
                                                 {\syncomposep{\transformer{2}}
                                                              {\transformer{2}'}}}}}}
          {\synvartermi{5}}
\\
\synappfour{\evaltwo{\synunit{3}}}{k_1}{k_2}{\synvartermi{5}}
& = &
\synpair{\synlam{\synvartermi{3}}{\synvartermi{3}}}
        {\synvartermi{5}}
\\
\synappfour{\evaltwo{\synprodp{3}{\synvarterm}{\synvarterm'}}}{k_1}{k_2}{\synvartermi{5}}
& = &
\vsynlet{\synpair{\transformer{3}}
                 {\synvartermi{5}}}
        {\synappfour{\evaltwo{\synvarterm}}
                    {k_1}
                    {k_2}
                    {\synvartermi{5}}}
        {\vsynlet{\synpair{\transformer{3}'}
                          {\synvartermi{5}}}
                 {\synappfour{\evaltwo{\synvarterm'}}
                             {k_1}
                             {k_2}
                             {\synvartermi{5}}}
                 {\synpair{\syncompose{\transformer{3}}
                                      {\transformer{3}'}}
                          {\synvartermi{5}}}}
\end{array}
\]

\[
\begin{array}[t]{@{}r@{\hspace{2mm}}c@{\hspace{2mm}}l@{}}
\synappfour{\evaltwo{\synunit{4}}}{k_1}{k_2}{\synvartermi{5}}
& = &
\synshiftn{1}
          {k_3}
          {\synpair{\synlam{\synvartermi{4}}{\synvartermi{4}}}
                   {\synvartermi{5}}}
\\
\synappfour{\evaltwo{\synprodp{4}{\synvarterm}{\synvarterm'}}}{k_1}{k_2}{\synvartermi{5}}
& = &
\synshiftn{1}
          {k_3}
          {\vsynlet{\synpair{\transformer{4}}
                            {\synvartermi{5}}}
                   {\synresetn{1}
                              {\synapp{k_3}
                                      {\synappfourp{\evaltwo{\synvarterm}}
                                                            {k_1}
                                                            {k_2}
                                                            {\synvartermi{5}}}}}
                   {\vsynlet{\synpair{\transformer{4}'}
                                     {\synvartermi{5}}}
                            {\synresetn{1}
                                       {\synapp{k_3}
                                               {\synappfourp{\evaltwo{\synvarterm'}}
                                                                     {k_1}
                                                                     {k_2}
                                                                     {\synvartermi{5}}}}}
                            {\synpair{\syncompose{\transformer{4}}
                                                 {\transformer{4}'}}
                                     {\synvartermi{5}}}}}
\\
\synappfour{\evaltwo{\synunit{5}}}{k_1}{k_2}{\synvartermi{5}}
& = &
\synshiftn{2}
          {k_4}
          {\synvartermi{5}}
\\
\synappfour{\evaltwo{\synprodp{5}{\synvarterm}{\synvarterm'}}}{k_1}{k_2}{\synvartermi{5}}
& = &
\synshiftn{1}
          {k_3}
          {\synshiftn{2}
                     {k_4}
                     {\vsynlet{\synvartermi{5}}
                              {\synresetn{2}
                                         {\synapp{k_4}
                                                 {\synresetn{1}
                                                            {\synapp{k_3}
                                                                    {\synappfourp{\evaltwo{\synvarterm'}}
                                                                                 {k_1}
                                                                                 {k_2}
                                                                                 {\synvartermi{5}}}}}}}
                              {\synresetn{2}
                                         {\synapp{k_4}
                                                 {\synresetn{1}
                                                            {\synapp{k_3}
                                                                    {\synappfourp{\evaltwo{\synvarterm}}
                                                                                 {k_1}
                                                                                 {k_2}
                                                                                 {\synvartermi{5}}}}}}}}}
\\[0.7cm]
\reifyn{2}{v}
& = &
\synresetnvlet{2}
              {\synpair{\transformer{4}}
                       {\synvartermi{5}}}
              {\synresetnvlet{1}
                             {\synpair{\transformer{3}}
                                      {\synvartermi{5}}}
                             {\synappfourv{v}
                                          {\synlamp{\transformer{1}}
                                                   {\synlam{k_2}
                                                           {\synapp{k_2}
                                                                   {\synlamp{\synvartermi{2}}
                                                                            {\dnfprod{2}
                                                                                     {\synappp{\transformer{1}}
                                                                                              {\dnfunit{1}}}
                                                                                     {\synvartermi{2}}}}}}}
                                          {\synlamp{\transformer{2}}
                                                   {\synlam{\synvartermi{3}}
                                                           {\dnfprod{3}
                                                                    {\synappp{\transformer{2}}
                                                                             {\dnfunit{2}}}
                                                                    {\synvartermi{3}}}}}
                                          {\synunit{5}}}
                             {\synpair{\synlam{\transformer{4}}
                                              {\dnfprod{4}
                                                       {\synappp{\transformer{3}}
                                                                {\dnfunit{3}}}
                                                       {\synvartermi{4}}}}
                                      {\synvartermi{5}}}}
              {\dnfprod{5}
                       {\synappp{\transformer{4}}
                                {\dnfunit{4}}}
                       {\synvartermi{5}}}
\\[2.1cm]
\norm{\synvarterm}
& = &
\reifyn{2}{\evaltwop{\synvarterm}}
\end{array}
\]

\noindent
Whereas $\rawevalfour$ had four layered continuations, $\rawevaltwo$ has
only two layered continuations since it has been mapped back to direct
style twice.  Where $\rawevalfour$ accesses $k_3$ as one of its
parameters, $\rawevaltwo$ abstracts the first layer of control with
shift$_1$, and where $\rawevalfour$ accesses $k_4$ as one of its
parameters, $\rawevaltwo$ abstracts the first and the second layer of
control with shift$_2$.

Uncurrying $k_1$ and $k_2$ and mapping $\rawevaltwo$ and $\rawreify_2$
back to direct style twice yield the following direct-style
definition, which lives at level 4 of the CPS hierarchy:
\[
\begin{array}{@{}r@{\hspace{2mm}}c@{\hspace{2mm}}l@{}}
\synapp{\evalzero{\synvar{x}}}{\synvartermi{5}}
& = &
\synpair{\synlam{\synvartermi{1}}{\dnfprod{1}{\synvar{x}}{\synvartermi{1}}}}
        {\synvartermi{5}}
\\
\synapp{\evalzero{\synunit{1}}}{\synvartermi{5}}
& = &
\synpair{\synlam{\synvartermi{1}}{\synvartermi{1}}}
        {\synvartermi{5}}
\\
\synapp{\evalzero{\synprodp{1}{\synvarterm}{\synvarterm'}}}{\synvartermi{5}}
& = &
\vsynlet{\synpair{\transformer{1}}
                 {\synvartermi{5}}}
        {\synapp{\evalzero{\synvarterm}}
                {\synvartermi{5}}}
        {\vsynlet{\synpair{\transformer{1}'}
                          {\synvartermi{5}}}
                 {\synapp{\evalzero{\synvarterm'}}
                         {\synvartermi{5}}}
                 {\synpair{\syncompose{\transformer{1}}
                                      {\transformer{1}'}}
                          {\synvartermi{5}}}}
\\
\synapp{\evalzero{\synunit{2}}}{\synvartermi{5}}
& = &
\synshiftn{1}
          {k_1}
          {\synpair{\synlam{\synvartermi{2}}{\synvartermi{2}}}
                   {\synvartermi{5}}}
\\
\synapp{\evalzero{\synprodp{2}{\synvarterm}{\synvarterm'}}}{\synvartermi{5}}
& = &
\synshiftn{1}
          {k_1}
          {\vsynlet{\synpair{\transformer{2}}
                            {\synvartermi{5}}}
                   {\synresetn{1}
                              {\synapp{k_1}
                                      {\synappp{\evalzero{\synvarterm}}
                                               {\synvartermi{5}}}}}
                   {\vsynlet{\synpair{\transformer{2}'}
                                     {\synvartermi{5}}}
                            {\synresetn{1}
                                       {\synapp{k_1}
                                               {\synappp{\evalzero{\synvarterm'}}
                                                        {\synvartermi{5}}}}}
                            {\synpair{\syncompose{\transformer{2}}
                                                 {\transformer{2}'}}
                                     {\synvartermi{5}}}}}
\\
\synapp{\evalzero{\synunit{3}}}{\synvartermi{5}}
& = &
\synshiftn{2}
          {k_2}
          {\synpair{\synlam{\synvartermi{3}}{\synvartermi{3}}}
                   {\synvartermi{5}}}
\\
\synapp{\evalzero{\synprodp{3}{\synvarterm}{\synvarterm'}}}{\synvartermi{5}}
& = &
\synshiftn{1}
          {k_1}
          {\synshiftn{2}
                     {k_2}
                     {\vsynlet{\synpair{\transformer{3}}
                                       {\synvartermi{5}}}
                              {\synresetn{2}
                                         {\synapp{k_2}
                                                 {\synresetn{1}
                                                            {\synapp{k_1}
                                                                    {\synappp{\evalzero{\synvarterm}}
                                                                             {\synvartermi{5}}}}}}}
                              {\vsynlet{\synpair{\transformer{3}'}
                                                {\synvartermi{5}}}
                                       {\synresetn{2}
                                                  {\synapp{k_2}
                                                          {\synresetn{1}
                                                                     {\synapp{k_1}
                                                                             {\synappp{\evalzero{\synvarterm'}}
                                                                                      {\synvartermi{5}}}}}}}
                                       {\synpair{\syncompose{\transformer{3}}
                                                            {\transformer{3}'}}
                                                {\synvartermi{5}}}}}}
\\
\synapp{\evalzero{\synunit{4}}}{\synvartermi{5}}
& = &
\synshiftn{3}
          {k_3}
          {\synpair{\synlam{\synvartermi{4}}{\synvartermi{4}}}
                   {\synvartermi{5}}}
\\
\synapp{\evalzero{\synprodp{4}{\synvarterm}{\synvarterm'}}}{\synvartermi{5}}
& = &
\synshiftn{1}
          {k_1}
          {\synshiftn{2}
                     {k_2}
                     {\synshiftn{3}
                                {k_3}
                                {\vsynlet{\synpair{\transformer{4}}
                                                  {\synvartermi{5}}}
                                         {\synresetn{3}
                                                    {\synapp{k_3}
                                                            {\synresetn{2}
                                                                       {\synapp{k_2}
                                                                               {\synresetn{1}
                                                                                          {\synapp{k_1}
                                                                                                  {\synappp{\evalzero{\synvarterm}}
                                                                                                           {\synvartermi{5}}}}}}}}}
                                         {\vsynlet{\synpair{\transformer{4}'}
                                                           {\synvartermi{5}}}
                                                  {\synresetn{3}
                                                             {\synapp{k_3}
                                                                     {\synresetn{2}
                                                                                {\synapp{k_2}
                                                                                        {\synresetn{1}
                                                                                                   {\synapp{k_1}
                                                                                                           {\synappp{\evalzero{\synvarterm'}}
                                                                                                                    {\synvartermi{5}}}}}}}}}
                                                  {\synpair{\syncompose{\transformer{4}}
                                                                       {\transformer{4}'}}
                                                           {\synvartermi{5}}}}}}}
\\
\synapp{\evalzero{\synunit{5}}}{\synvartermi{5}}
& = &
\synshiftn{4}
          {k_4}
          {\synvartermi{5}}
\\
\synapp{\evalzero{\synprodp{5}{\synvarterm}{\synvarterm'}}}{\synvartermi{5}}
& = &
\synshiftn{1}
          {k_1}
          {\synshiftn{2}
                     {k_2}
                     {\synshiftn{3}
                                {k_3}
                                {\synshiftn{4}
                                           {k_4}
                                           {\vsynlet{\synvartermi{5}}
                                                    {\synresetn{4}
                                                               {\synapp{k_4}
                                                                       {\synresetn{3}
                                                                                  {\synapp{k_3}
                                                                                          {\synresetn{2}
                                                                                                     {\synapp{k_2}
                                                                                                             {\synresetn{1}
                                                                                                                        {\synapp{k_1}
                                                                                                                                {\synappp{\evalzero{\synvarterm'}}
                                                                                                                                         {\synvartermi{5}}}}}}}}}}}
                                                    {\synresetn{4}
                                                               {\synapp{k_4}
                                                                       {\synresetn{3}
                                                                                  {\synapp{k_3}
                                                                                          {\synresetn{2}
                                                                                                     {\synapp{k_2}
                                                                                                             {\synresetn{1}
                                                                                                                        {\synapp{k_1}
                                                                                                                                {\synappp{\evalzero{\synvarterm}}
                                                                                                                                         {\synvartermi{5}}}}}}}}}}}}}}}

   \end{array}
 \]
 \[
 \begin{array}[t]{r@{\hspace{2mm}}c@{\hspace{2mm}}l}
\reifyn{4}{v}
& = &
\synresetnvlet{4}
              {\synpair{\transformer{4}}
                       {\synvartermi{5}}}
              {\synresetnvlet{3}
                             {\synpair{\transformer{3}}
                                      {\synvartermi{5}}}
                             {\synresetnvlet{2}
                                            {\synpair{\transformer{2}}
                                                     {\synvartermi{5}}}
                                            {\synresetnvlet{1}
                                                           {\synpair{\transformer{1}}
                                                                    {\synvartermi{5}}}
                                                           {\synapp{v}
                                                                   {\synunit{5}}}
                                                           {\synpair{\synlam{\transformer{2}}
                                                                            {\dnfprod{2}
                                                                                     {\synappp{\transformer{1}}
                                                                                              {\dnfunit{1}}}
                                                                                     {\synvartermi{2}}}}
                                                                    {\synvartermi{5}}}}
                                            {\synpair{\synlam{\transformer{3}}
                                                             {\dnfprod{3}
                                                                      {\synappp{\transformer{2}}
                                                                               {\dnfunit{2}}}
                                                                      {\synvartermi{3}}}}
                                                     {\synvartermi{5}}}}
                             {\synpair{\synlam{\transformer{4}}
                                              {\dnfprod{4}
                                                       {\synappp{\transformer{3}}
                                                                {\dnfunit{3}}}
                                                       {\synvartermi{4}}}}
                                      {\synvartermi{5}}}}
              {\dnfprod{5}
                       {\synappp{\transformer{4}}
                                {\dnfunit{4}}}
                       {\synvartermi{5}}}
\\[2.1cm]
\norm{\synvarterm}
& = &
\reifyn{4}{\evalzerop{\synvarterm}}
\end{array}
\]

\noindent
Whereas $\rawevaltwo$ had two layered continuations, $\rawevalzero$ has
none since it has been mapped back to direct style twice.  Where
$\rawevaltwo$ accesses $k_1$ as one of its parameters, $\rawevalzero$
abstracts the first layer of control with shift$_1$, and where
$\rawevaltwo$ accesses $k_2$ as one of its parameters, $\rawevalzero$
abstracts the first and the second layer of control with shift$_2$.
Where $\rawevaltwo$ uses reset$_1$ and shift$_1$, $\rawevalzero$ uses
reset$_3$ and shift$_3$, and where $\rawevaltwo$ uses reset$_2$ and
shift$_2$, $\rawevalzero$ uses reset$_4$ and shift$_4$.

\subsection{A note about efficiency}

We have implemented all the definitions of
Section~\ref{subsec:a-generalization-of-propositional-logic} as well as
the intermediate versions $\mathit{eval}_1$ and $\mathit{eval}_3$
in ML~\cite{Danvy-Yang:ESOP99}.  We have also
implemented hierarchical normalization functions for other values than 5.

For high products (\ie, in
Section~\ref{subsec:a-generalization-of-propositional-logic}, for source
terms using $\rawsynprod{3}$ and $\rawsynprod{4}$), the normalization
function living at level 0 of the CPS hierarchy is the most efficient
one.  On the other hand, for low products (\ie, in
Section~\ref{subsec:a-generalization-of-propositional-logic}, for source
terms using $\rawsynprod{1}$ and $\rawsynprod{2}$), the normalization
functions living at
a higher level of the CPS hierarchy are the most efficient ones.  These
relative efficiencies are explained in terms of resources:
\begin{itemize}
  
\item Accessing to a continuation as an explicit parameter is more
  efficient than accessing to it through a control operator.
  
\item On the other hand, the restriction of $\rawevalzero$ to source
  terms that only use $\synunit{1}$ and $\rawsynprod{1}$ is in direct
  style, whereas the corresponding restrictions of $\rawevaltwo$ and
  $\rawevalfour$ pass a number of extra parameters.  These extra
  parameters penalize performance.

\end{itemize}

\noindent
The better performance of programs in the CPS hierarchy has already been
reported for level~1 in the context of continuation-based partial
evaluation~\cite{Lawall-Danvy:LFP94}, and it has been reported for a
similar ``pay as you go'' reason: a program that abstracts control
relatively rarely is run more efficiently in direct style with a control
operator rather than in continuation-passing style.

\subsection{Summary and conclusion}

We have illustrated the CPS hierarchy with an application of
normalization by evaluation that naturally involves successive layers of
continuations and that demonstrates the expressive power of shift$_n$ and
reset$_n$.

The application also suggests alternative control operators that would
fit better its continu\-ation-based programming pattern.  For example,
instead of representing a delimited continuation as a function and apply
it as such, we could represent it as a continuation and apply it with
a `throw' operator
as in MacLisp
and Standard ML of New Jersey.  For another example, instead of throwing
a value to a continuation, we could specify the continuation of a
computation, \eg, with a $\synvar{reflect}_i$ special form.  For a third
example, instead of abstracting control up to a layer $n$, we could give
access to each of the successive layers up to $n$, \eg, with a ${\mathcal
L}_n$
operator.  Then instead of
\[
\begin{array}[t]{@{\hspace{5mm}}r@{\hspace{2mm}}c@{\hspace{2mm}}l}
\synapp{\evalzero{\synprodp{4}{\synvarterm}{\synvarterm'}}}{\synvartermi{5}}
& = &
\synshiftn{1}
          {k_1}
          {\synshiftn{2}
                     {k_2}
                     {\synshiftn{3}
                                {k_3}
                                {\vsynlet{\synpair{\transformer{4}}
                                                  {\synvartermi{5}}}
                                         {\synresetn{3}
                                                    {\synapp{k_3}
                                                            {\synresetn{2}
                                                                       {\synapp{k_2}
                                                                               {\synresetn{1}
                                                                                          {\synapp{k_1}
                                                                                                  {\synappp{\evalzero{\synvarterm}}
                                                                                                           {\mbox{\hspace{0.2mm}}\synvartermi{5}}}}}}}}}
                                         {\vsynlet{\synpair{\transformer{4}'}
                                                           {\synvartermi{5}}}
                                                  {\synresetn{3}
                                                             {\synapp{k_3}
                                                                     {\synresetn{2}
                                                                                {\synapp{k_2}
                                                                                        {\synresetn{1}
                                                                                                   {\synapp{k_1}
                                                                                                           {\synappp{\evalzero{\synvarterm'}}
                                                                                                                    {\synvartermi{5}}}}}}}}}
                                                  {\synpair{\syncompose{\transformer{4}}
                                                                       {\transformer{4}'}}
                                                           {\synvartermi{5}}}}}}}
\end{array}
\]

\noindent
one could write
\[
\begin{array}[t]{@{\hspace{6mm}}r@{\hspace{2mm}}c@{\hspace{2mm}}l}
\synapp{\evalzero{\synprodp{4}{\synvarterm}{\synvarterm'}}}{\synvartermi{5}}
& = &
{\mathcal L}_3\:\syntriple{k_1}{k_2}{k_3}.\vsynlet{\synpair{\transformer{4}'}
                                                  {\synvartermi{5}}}
                                         {\synapp{\synvar{reflect}_3}
                                                 {\synquadruple{\synappthree{\synvar{eval}_4}
                                                                            {\synvarterm}
                                                                            {\mbox{\hspace{0.2mm}}\synvartermi{5}}}
                                                               {k_1}
                                                               {k_2}
                                                               {k_3}}}
                                         {\vsynlet{\synpair{\transformer{4}'}
                                                           {\synvartermi{5}}}
                                                  {\synapp{\synvar{reflect}_3}
                                                          {\synquadruple{\synappthree{\synvar{eval}_4}
                                                                                     {\synvarterm'}
                                                                                     {\synvartermi{5}}}
                                                                        {k_1}
                                                                        {k_2}
                                                                        {k_3}}}
                                                  {\synpair{\syncompose{\transformer{4}}
                                                                       {\transformer{4}'}}
                                                           {\synvartermi{5}}.}}
\end{array}
\]

\noindent
Such alternative control operators can be more convenient to use, while
being compatible with CPS.

\section{Conclusion and issues}
\label{sec:concl}

\noindent
We have used CPS as a guideline to establish an operational foundation
for delimited continuations.  Starting from a call-by-value evaluator
for $\lambda$-terms with shift and reset, we have mechanically
derived the corresponding abstract machine.  From this abstract
machine, it is straightforward to
obtain a reduction semantics of delimited control that, by construction,
is compatible with CPS---both for one-step reduction and for evaluation.
These results can also be established without the guideline of CPS, but less
easily.

\myindent
The whole approach
generalizes straightforwardly to account for the shift$_n$ and
reset$_n$ family of delimited-control operators and more generally for
any control operators that are compatible with CPS.  These results would
be non-trivial to establish without the guideline of CPS.

\myindent
Defunctionalization provides a key for connecting
continuation-passing style and operational intuitions about control.
Indeed most of the time, control stacks and evaluation contexts are
the defunctionalized continuations of an evaluator.
Defunctionalization also provides a key for identifying where
operational intuitions about control go beyond CPS (see
Section~\ref{subsec:beyond-CPS}).  

\myindent
We do not know whether CPS is the
ultimate answer, but the present work shows yet another example of its
usefulness.  It is like nothing can go wrong with CPS.
\section*{Acknowledgments}
We are grateful to Mads Sig Ager, Julia Lawall, Jan Midtgaard, and the
referees of CW'04 and of LMCS for their comments.  The third author would
also like to thank Samuel Lindley for our joint initial study of the
normalization functions of Section~\ref{sec:programming-in-hierarchy}.

This work is partially supported by the ESPRIT Working Group APPSEM~II\break
({\small\url{http://www.appsem.org}}), by the Danish Natural Science
Research Council, Grant no.~21-02-0474 (for the two first authors) and
Grant no.~21-03-0545 (for the third author), and by BRICS (Basic Research
in Computer Science ({\small\url{http://www.brics.dk}}), funded by the
Danish National Research Foundation).

\bibliography{biernacka-biernacki-danvy-LMCS}
\bibliographystyle{plain}

\end{document}